\documentclass[twocolumn]{RAA}     
   
\usepackage{graphicx,times}
\usepackage{natbib}
\usepackage{amssymb,amsmath}
\bibpunct{(}{)}{;}{a}{}{,}
\graphicspath{{./}{figures/}}

\defcitealias{Ju2022}{Ju22}
\defcitealias{BC03}{BC03}

\newcommand{\oiii}{\hbox{[O\,{\scriptsize III}]}}
\newcommand{\nii}{\hbox{[N\,{\scriptsize II}]}}

\newcommand{\Hii}{\hbox{H\,{\scriptsize II}}}                                                       
\newcommand{\Hi}{\hbox{H\,{\scriptsize I}}}

\usepackage[pagebackref=true]{hyperref}

\begin{document}

   \title{The Clumpy Structure Of Five Star-bursting Dwarf Galaxies In The MaNGA Survey}

 \volnopage{ {\bf 20XX} Vol.\ {\bf X} No. {\bf XX}, 000--000}
   \setcounter{page}{1}

   \author{Mengting Ju 
      \inst{1,2}
   \and Jun Yin
      \inst{1,3}
   \and Lei Hao
      \inst{1}
    \and Chenxu Liu
      \inst{4}
   \and Chao-Wei Tsai
      \inst{5,6,2}
   \and Junfeng Wang
      \inst{7}
    \and Zhengyi Shao
      \inst{1,3}
    \and Shuai Feng
      \inst{8,9}
    \and Yu Rong
      \inst{10}
   }

   \institute{Key Laboratory for Research in Galaxies and Cosmology, Shanghai Astronomical Observatory, Chinese Academy of Sciences, 80 Nandan Road, Shanghai 200030, People's Republic of China; {\it mtju@shao.ac.cn, jyin@shao.ac.cn, haol@shao.ac.cn}\\
        \and
             University of Chinese Academy of Sciences, 19A Yuquan Road, Beijing 100049, People's Republic of China\\
        \and
             Key Lab for Astrophysics, Shanghai, 200034, People's Republic of China\\
        \and
             South-Western Institute for Astronomy Research, Yunnan University, Kunming, Yunnan, 650500, People's Republic of China\\
        \and
             National Astronomical Observatories, Chinese Academy of Sciences, 20A Datun Road, Beijing 100101, People's Republic of China\\   
         \and
             Institute for Frontiers in Astronomy and Astrophysics, Beijing Normal University, Beijing 102206, People's Republic of China\\  
        \and
             Department of Astronomy, Physics Building, Xiamen University, Xiamen, Fujian, 361005, People's Republic of China\\ 
        \and
            College of Physics, Hebei Normal University, 20 South Erhuan Road, Shijiazhuang 050024, People's Republic of China\\
        \and
            Hebei Key Laboratory of Photophysics Research and Application, Shijiazhuang 050024, People's Republic of China\\
        \and
            Department of Astronomy, University of Science and Technology of China, No.96, JinZhai Road, Baohe District, Hefei, Anhui, 230026, People's Republic of China\\
\vs \no
   {\small Received 20XX Month Day; accepted 20XX Month Day}
}

\abstract{The star-forming clumps in star-bursting dwarf galaxies provide valuable insights into the understanding of the evolution of dwarf galaxies. In this paper, we focus on five star-bursting dwarf galaxies featuring off-centered clumps in the Mapping Nearby Galaxies at Apache Point Observatory (MaNGA) survey. Using the stellar population synthesis software {\tt FADO}, we obtain the spatially-resolved distribution of the star formation history, which allows us to construct the $g$-band images of the five galaxies at different ages. These images can help us to probe the evolution of the morphological structures of these galaxies. While images of stellar population older than 1~Gyr are typically smooth, images of stellar population younger than 1~Gyr reveal significant clumps, including multiple clumps which appear at different locations and even different ages.
To study the evolutionary connections of these five galaxies to other dwarf galaxies before their star-forming clumps appear, we construct the images of the stellar populations older than three age nodes, and define them to be the images of the ``host" galaxies. We find that the properties such as the central surface brightness and the effective radii of the hosts of the five galaxies are in between those of dwarf ellipticals (dEs) and dwarf irregulars (dIrrs), with two clearly more similar to dEs and one more similar to dIrrs. Among the five galaxies, 8257-3704 is particularly interesting, as it shows a previous starburst event that is not quite visible from its $gri$ image, but only visible from images of the stellar population at a few hundred million years. The star-forming clump associated with this event may have appeared at around 600~Myr and disappeared at around 40~Myr.
}
   \authorrunning{Mengting Ju et al. }            
   \titlerunning{Off-centered Clumps in Star-bursting Dwarf Galaxies }  

   \maketitle

%
\section{Introduction}           
\label{sect:intro}

Dwarf galaxies are the most abundant structures in the universe, and their unique properties offer crucial insights into fundamental astrophysical processes and a broader understanding of galaxy formation and evolution. Star-bursting dwarf galaxies, such as the Blue Compact Dwarf (BCD), are a type of dwarf galaxies with high star formation rates (SFRs). Some star-bursting dwarf galaxies have giant star formation clumps \citep{Elmegreen2012a, Verbeke2014, Zhang2020}. These galaxies are usually gas-rich and tend to have low metallicities \citep{Hunter1999, Zhang2012}. Some of them have extremely low metallicities, similar to Population I systems. It is proposed that they are very young with the first generation of star formation or the star formation in these galaxies occurs in intense bursts separated by extended quiescent periods \citep{Searle1972}. 
Several works confirmed that there are old stellar populations in star-bursting dwarf galaxies \citep{Cair2001, Cair2003, Annibali2013}. This leads us to speculate whether these star-bursting dwarf galaxies, when in the quiescent stage, are connected to other types of dwarf galaxies. Some people have attempted to use galaxy images to investigate these connections \citep[e.g.][]{Papaderos1996, Papaderos2008, Amorin2009, Meyer2014, Lian2015, Rey2023}, by comparing the properties of the underlying old-stellar-population hosts with other dwarf galaxies. For example, some studies \citep[e.g.][]{Amorin2009, Lian2015} have reported that the effective radii of the hosts of the starbursting dwarfs are smaller than those of dwarf irregular galaxies (dIrrs), although such results have not been consistently confirmed by studies using the near-infrared images. These studies suffer a substantial challenge in separating the active star formation clumps from the hosts. If the analysis is done on the optical images where the star-forming clumps can be significantly visible, only the regions where the contributions from the star-bursting regions are minimal can be used. This puts a strong limitation on this area of study. 

The star formation clumps found in local star-bursting dwarf galaxies, even though they can be contaminants in studies of the hosts, are by themselves valuable in understanding the evolution of these galaxies due to their resemblance to the clumpy structures observed in high-redshift galaxies. High-redshift galaxies are known to be clumpy \citep{Elmegreen2004, Conselice2005}. The clumps in high-redshift and local galaxies show great similarities, both exhibiting high SFR and low gas-phase metallicity\citep{Erb2006, Almeida2014, Almeida2018, Lagos2018}, except that clumps in high-redshift galaxies tend to be larger in size compared to those in local galaxies \citep{Genzel2006, Elmegreen2013}. In \citet[][hereafter \citetalias{Ju2022}]{Ju2022}, we reported a distinct off-centered clump in the star-bursting dwarf galaxy (8313-1901) in the Mapping Nearby Galaxies at Apache Point Observatory survey \citep[MaNGA;][]{bundy2015}. The clump has a size comparable to the clumps at high redshifts. By analysing the SFR, metallicity, kinematics, and particularly the stellar populations of the clump, as well as those of the host, we concluded that this clump likely originated from a gas accretion event. 

The primary analysis of the clump in 8313-1901 by \citetalias{Ju2022} was done on the Integral Field Spectroscopy (IFS) data. Unlike traditional aperture or long slit spectrographs, which provide light within a small region of galaxies, IFS allows simultaneous spectroscopic observations of multiple regions across the field of view. The spatially-resolved spectra can be used to separate the physical properties, such as gas-phase metallicity, stellar populations and star formation history, in the clump regions and other locations.

Inspired by the analysing of the clump in the MaNGA 8313-1901, in this paper, we investigate five MaNGA galaxies, including 8313-1901, selected by their dynamically incoherent star formation clumps with respect to the host galaxies. We study both their hosts and the clumps using a dedicated spectral analysis tool: the Fitting Analysis using Differential evolution Optimization \citep[{\tt FADO};][]{FADO2017}. The paper is structured as follows: In Section~\ref{sec:data}, we introduce the MaNGA survey data of the five galaxies. In Section~\ref{sec:results}, we analyze the physical properties of these five dwarfs and their stellar populations. In Section~\ref{sec:discussion}, we discuss the structural changes. A summary is presented in Section~\ref{sec:summary}. Throughout this paper, we adopt cosmological parameters of $H_0=70~\rm km/s/Mpc,\Omega_{M}=0.3$, and $\Omega_\Lambda=0.7$.

\section{Data}
\label{sec:data}

\begin{table*}
  \caption{Parameters of the five star-bursting dwarf galaxies.}
  \tabcolsep=0.10cm
  \label{table:1}
  \centering
  \begin{tabular}{l c c c c c c c c c }
\hline\hline
SDSS &plateifu  & RA (J2000) & Dec (J2000) & redshift$^a$ & $NUV-r^a$ & $M_g^a$ & ${\rm log(M_*)}^a$ & ${\rm log(M_{\Hi})^b}$  \\
  & & $^\circ$& $^\circ$  & & (mag) & (mag) & (${\rm M_{\odot}})$ & (${\rm M_{\odot}})$   \\
\hline
J110212.87+451813.7 & 8257-3704	& 165.5536	& 45.3039	& 0.0202 & 2.76 & -17.54 & 8.69 &9.36 \\
 J160108.90+415250.7& 8313-1901	& 240.2871	& 41.8807	& 0.0243 & 1.40 & -18.64 & 8.88  & 9.37  \\
J160820.11+494802.4 & 8563-3704	& 242.0838	& 49.8007	& 0.0192 & 1.70	& -17.57 & 8.63  & -  \\
J212417.32+010142.1 & 8615-1901	& 321.0722	& 1.0284	& 0.0197 & 1.30	& -18.59 & 8.77  & 9.64  \\
J164625.11+194611.5 & 9894-9102	& 251.6052	& 19.7696	& 0.0232 & 1.99	& -18.59 & 8.79  & 9.76  \\

\hline
\end{tabular}\\
Notes:
$^a$ the NASA-Sloan Atlas catalog: http://www.nsatlas.org\\
$^{b}$ \cite{Masters2019,Stark2021}\\

\end{table*}

\begin{figure*}
    \centering
    \includegraphics[width=0.90\textwidth,clip,trim={0 0 0 0}]{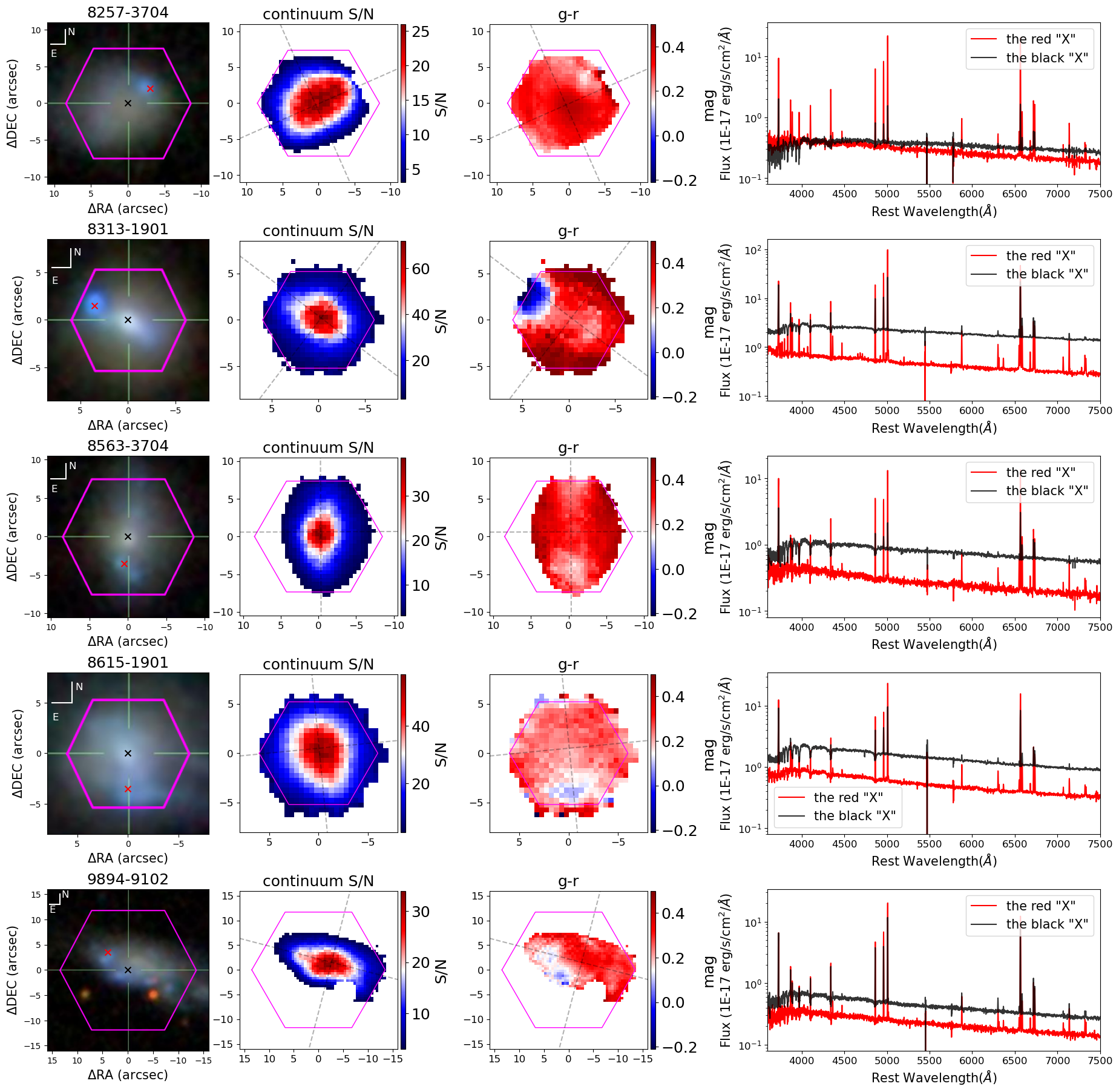} 
    \caption{MaNGA plate IFU ID from top to bottom: 8257-3704, 8313-1901, 8563-3704, 8615-1901, 9894-9102. From left to right: SDSS $gri$ composite images, the mean signal-to-noise ratio (S/N) of the continuum between 3600~\AA and 7500~\AA, $g-r$ color, and two spaxel spectra that are located at the red and black crosses marked in the $gri$ composite images. The magenta hexagons in the first three columns show the coverage of the MaNGA bundle. We only show the spaxels with S/N greater than 3. The gray dashed lines in the second and third columns indicate the major and minor axes of each galaxy, which are determined by fitting the two-dimensional surface brightness profiles of the host galaxies, as described in Section~\ref{sec:hosts}. In the rest two-dimensional images presented in this paper, unless otherwise specified, the gray dashed lines represent the major and minor axes of the galaxies. The black curves in the fourth column represent the spectra from the center of the MaNGA FoV, and the red curves show the spectra in the clumps.  }
    \label{fig:image}
\end{figure*}

The MaNGA survey observed approximately 10,000 nearby galaxies ($<z>\ \sim 0.03$). Among them, there are nearly 1,500 dwarf galaxies with stellar masses less than $10^9 M_{\odot}$. We tentatively compile a sample of BCD candidates from the MaNGA survey using the BCD criteria outlined in \cite{gil2003} and \cite{Sanchez2008}. The criteria are set to be that these galaxies are blue ($\left \langle \mu_B \right \rangle-\left \langle \mu_R \right \rangle \leq 1\  mag/arcsec^2$), compact ($\left \langle \mu_B \right \rangle < 22\ mag/arcsec^2$), and dwarfy ( $M_{*} < 10^9 M_{\odot}$). Based on the measurements in the NASA-Sloan Atlas catalog, we identify 53 galaxies that meet these criteria. Approximately half of them exhibit off-center clumps, and among them we select five representative star-bursting dwarf galaxies. Their $gri$ composite images from SDSS are shown in the first column of Figure~\ref{fig:image}. 

MaNGA galaxy 8257-3704 is characterized by a distinct clump to the northwest. 8313-1901 features a significant clump to the northeast, which we investigated in detail in \citetalias{Ju2022}). 8563-3704 has two blue clumps: one to the north and the other to the south. 8615-1901 has a blue clump covering the entire galaxy. 9894-9102 has a sequence of multiple clumps that are close to each other.

Table~\ref{table:1} presents the coordinates, redshifts, $NUV-r$ colors, $g$-band absolute magnitudes, stellar masses, $\Hi$ gas masses, and environmental information of these galaxies. The coordinates, redshifts, $NUV-r$ colors, $g$-band absolute magnitudes, and the stellar mass are obtained from the NASA-Sloan Atlas catalog. The magnitudes, including $NUV, g$, and $r$-bands, and the stellar masses are the elliptical Petrosian parameters in the NASA-Sloan Atlas catalog. The $\Hi$ gas masses in Table~\ref{table:1} are obtained from the HI-MaNGA follow-up survey, which includes information from Green Bank Telescope (GBT) observations and from the Arecibo Legacy Fast ALFA (ALFALFA) project \citep{Masters2019,Stark2021}.


The MaNGA survey is an IFS survey and is a part of the fourth generation of the Sloan Digital Sky Survey \citep[SDSS-IV;][]{blanton2017}. It employs 17 science Integral Field Units (IFUs) ranging in size from 19 fibers to 127 fibers (12\arcsec\ - 32\arcsec\ diameter). 
The selected IFUs are required to cover the galaxies up to 1.5 and 2.5 effective radii ($R_e$) \citep{yan2016b}. The spectra are obtained with the Baryon Oscillation Spectroscopic Survey (BOSS) spectrographs \citep{smee13} on the 2.5-meter Sloan Telescope at Apache Point Observatory \citep{gunn06}. The typical MaNGA reduced data cubes have a spaxel size of 0.\arcsec5 and a point spread function (PSF) with a full width at half-maximum (FWHM) of 2.\arcsec5 \citep{Law2015, yan2016b}.
The MaNGA survey provides data covering a wavelength range from 3600\,\AA\ to 10,300\,\AA\ with a resolution of $R \sim\ 2000$ \citep{law2016, Law2021, yan2016a}. In this work, we mainly use the data product from the MaNGA Data Reduction Pipeline MPL-11 version \citep[{\tt{DRP}};][]{law2016}. The maps of the median S/N of the continuum averaged over the wavelength range of 3600\,\AA\ $< \lambda < $7500\,\AA\ is displayed in the second column of Figure~\ref{fig:image}. In the five galaxies, there are 3839 spaxels with a continuum S/N greater than 3, covering a significant fraction of the observed field (82\%). The spectra in the central regions have S/N of several tens. In this paper, we use only the 3839 spaxels for further analysis. 

These spectra suffer certain amount of extinction by ISM in our Milky Way. Thus, we correct all these observed spectra for the Milky Way foreground extinction law of \cite{Fitzpatrick1999} with the Galactic reddening $E(B-V)$ provided by the MaNGA database. We then de-redshift the spectra using the redshift values of the five galaxies provided by the NASA-Sloan Atlas catalog. We refer to these corrected spectra as the corrected-observed (Cobs) spectra. The third column of Figure~\ref{fig:image} shows the two-dimensional distribution map of the $g-r$ color synthesized from the Cobs spectra. The magnitudes in $g$- and $r$-band are calculated by convolving the Cobs spectra with the filter response curves. It can be seen that the clump regions are generally bluer than the other regions in galaxies which is consistent with the $gri$ composite images. In the fourth column, we present the Cobs spectra of two spaxels in each galaxy marked by the red cross and the black cross in the $gri$ composite images, respectively. The red cross marks a representative position in the clump, while the black cross represents the center of the MaNGA field-of-view (FoV).

\section{Results}
\label{sec:results}

\subsection{The Spectral Fitting Analyses}
\label{sec_ssp}

\begin{figure}
    \centering
    \includegraphics[width=0.5\textwidth,clip,trim={0 0 0 0}]{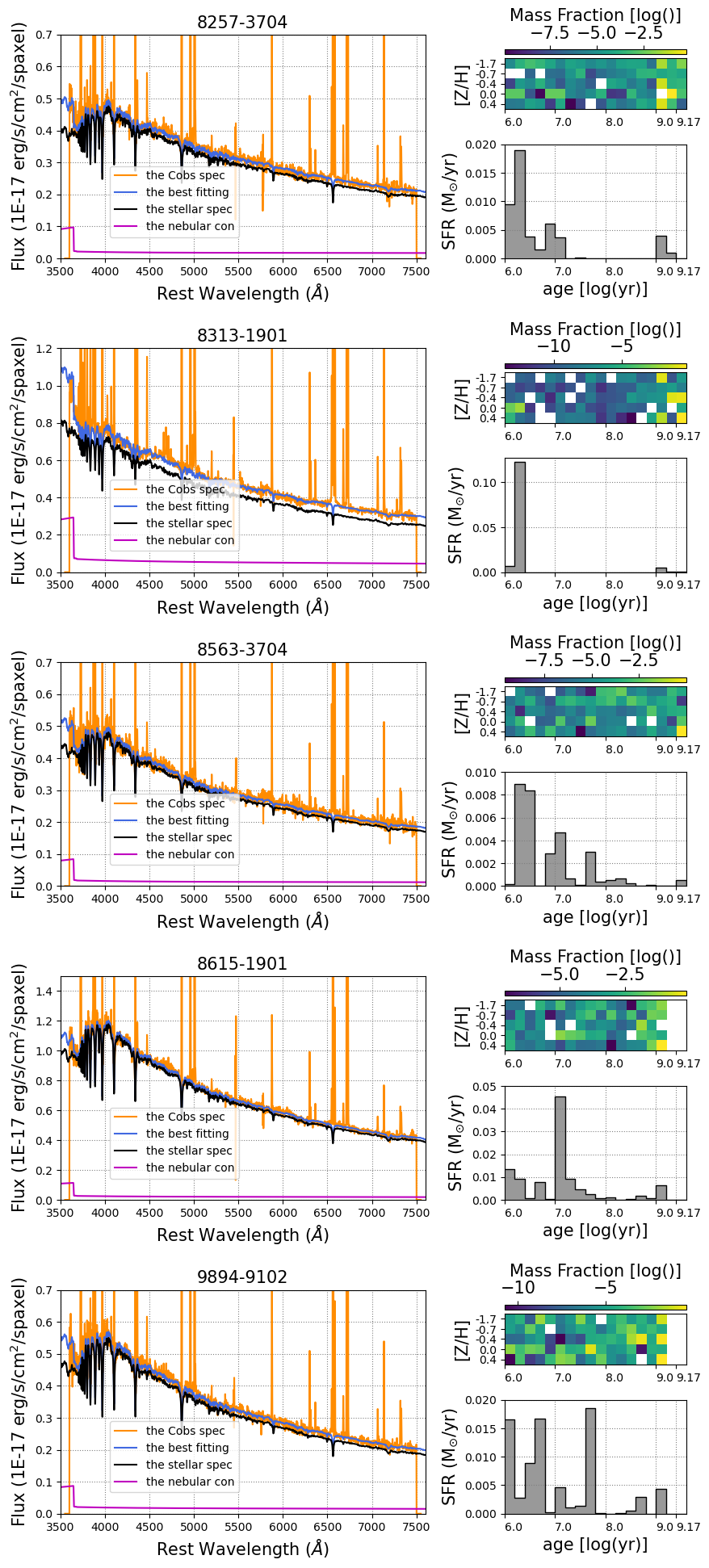} 
    \caption{Examples of the {\tt FADO} fitting results for the five galaxies. The orange lines in the left panels show the Cobs spectra in clump regions. The blue lines are the best-fit continuum. The black lines are the stellar continuum. The purple lines are the nebular continuum. 
    The right panels show the contributions of SSPs that form the best-fit populations in the age-metallicity maps (top) and the evolution of SFR (bottom), normalized by stellar mass.}
    \label{fig:hostspec}
\end{figure}

The MaNGA spectra of the five star-bursting galaxies are analysed with the Fitting Analysis using Differential evolution Optimization \citep[{\tt FADO};][]{FADO2017}. {\tt FADO} is a spectral stellar population synthesis tool. Unlike most currently available stellar population synthesis codes, the fitting of the stellar continuum in {\tt FADO} does not mask nebular emission lines. Instead, it incorporates self-consistent observed nebular emission (both nebular continuum and emission lines) that based on the star formation history (SFH) and chemical enrichment history inferred from the best stellar models. The nebula continuum is formed when the Lyman continuum (LyC) photons ($\lambda < $911.76\,\AA\ ) emitted by the stars are absorbed by the \Hii\ region gas and re-emitted at longer wavelengths. In FADO, the spectrum of the nebular continuum is included in the fitting, assuming case B recombination under typical physical conditions of the \Hii\ region, i.e., an electron density of $ne$ = 100 cm$^3$ and temperature of T$_e$ = 10$^4$ K. 
Nebular emission can significantly contaminate the host continuum photometry in the star-forming galaxies, and further affect spectral energy distribution (SED) studies \citep{Salzer1989, Jaskot2013}. It is crucial to remove the contribution of nebular emission from the stellar continuum SED of star-forming regions; otherwise old stellar populations might be overestimated \citep{Izotov2011}. \cite{Pappalardo2021} compared the results of {\tt FADO} with two other spectral stellar population synthesis models, on a set of simulated galaxy spectra which were constructed as having both the stellar and nebular emissions. For the two models that do not include the nebular emissions in the fitting, the mass-weighted mean age obtained is overestimated by $\sim$ 2 dex at a young stellar age, if the contributions of nebular emissions are not considered in the spectra.

We use {\tt FADO (v.1B)} to fit the 3839 Cobs spectra, covering the wavelength range from 3600\,\AA\ to 7500\,\AA\ and estimating the host galaxy extinction using the extinction law of \cite{Calzetti2001}. We choose single stellar populations (SSPs) from the \citet[][hereafter \citetalias{BC03}]{BC03} which were build on the Padova1994 stellar evolution tracks \citep{Fagotto1994a, Fagotto1994b} and the Salpeter initial mass function (IMF). A combination of 90 SSPs with 18 ages (age = 1.00~Myr, 1.58~Myr, 2.51~Myr, 3.98~Myr, 6.31~Myr, 10.00~Myr, 15.85~Myr, 25.12~Myr, 40.00~Myr, 64.05~Myr, 101.52~Myr, 160.90~Myr, 255.00~Myr, 404.15~Myr, 640.54~Myr, 1.02~Gyr, 3.75~Gyr, and 15.00~Gyr, ranged from 1Myr to 15 Gyr) and 5 metallicities (Z = 0.0004, 0.004, 0.008, 0.02, and 0.05) are used. 

Figure \ref{fig:hostspec} presents the {\tt FADO} fitting results for the five galaxies. From top to bottom, we show the best-fit spectrum in a representative spaxel in the clump region of 8257-3704, 8313-1901, 8563-3704, 8615-1901, and 9894-9102, respectively. The orange curve represents the Cobs spectrum. The best-fit continuum is shown in the blue. The continuum consists of the stellar continuum (black) and the nebular continuum (purple). The right panels of Figure~\ref{fig:hostspec} show the fraction of stellar mass of each SSP that constitute the best-fitting populations in the age-metallicity maps (the top panels) and in the SFH (the bottom panels). The stellar continua can be reconstructed with the SFH and the 90 SSPs.

\begin{figure*}
\centering
\includegraphics[width=1.0\textwidth,clip,trim={0 0 0 0}]{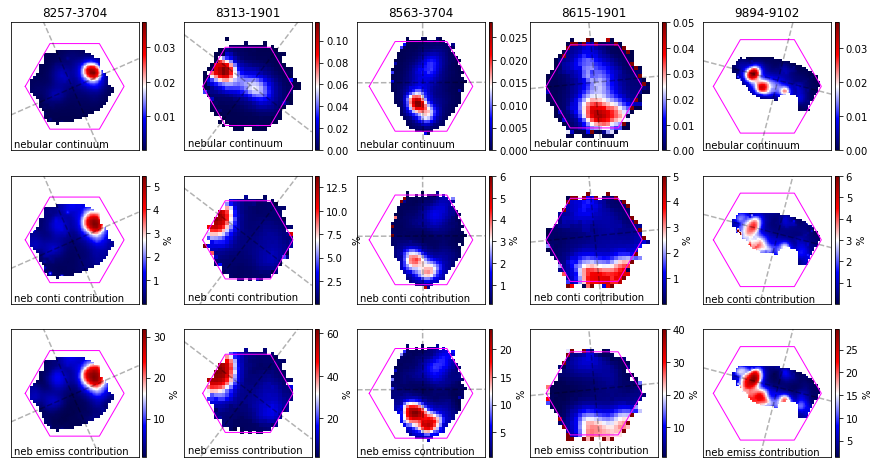} 
   \caption{Top row: the $g$-band flux of nebular continua. Middle row: the percentage of the flux of the nebular continua in the overall continua. Bottom row: the percentage of the flux of the nebular emission contributing to the Cobs spectra.}
    \label{fig:nebul}
\end{figure*}

In Figure~\ref{fig:hostspec}, we find that the nebular continuum in the best-fit result does have a non-zero contribution. In Figure~\ref{fig:nebul}, we investigate the nebular emission in detail, calculating the nebular contributions of each galaxy. In the top row of Figure~\ref{fig:nebul}, we show the $g$-band images of the nebular continua and the unit is nanomaggies. The middle row shows the ratio of the nebular continua to the overall continua in $g$-band flux. The nebular continua contributions are nearly 5\% in the clump regions in most of the five galaxies. 8313-1901, which exhibits the highest SFR in the five star-bursting galaxies, has a higher nebular contribution ($\sim$10\%) compared to other four galaxies. The bottom row in Figure~\ref{fig:nebul} shows the ratio of the $g$-band flux from the nebular emissions, including the nebular continuum and emission lines to the $g$-band flux from the Cobs spectra. The nebular emission contributions are significantly, from 30\% to 60\%, to the optical flux in the clump regions. This is consistent with the findings of other studies \citep[e.g.][]{Krueger1995}.

We check {\tt FADO} results with our previous findings in \citetalias{Ju2022}. In \citetalias{Ju2022}, we constructed the clump spectrum in 8313-1901. The clump spectrum is obtained by subtracting the observed spectrum from the host contribution. It matches very well with the young model spectrum ($\leq$ 7~Myr), which has stellar mass of log(M/M$_\odot$) = 6.26. 
We use {\tt FADO} and 90 SSP models to fit the clump spectrum. The {\tt FADO} results, including the stellar populations and stellar mass, are consistent with the results of \citetalias{Ju2022}. The stellar population is very young ($\sim$ 4~Myr) and the stellar mass is about log(M/M$_\odot$)=6.12.

\subsection{SFR, Gas-phase Metallicity and Kinematics}
\label{sec:method}

\begin{figure*}
    \centering
    \includegraphics[width=0.8\textwidth,clip,trim={0 0 0 0}]{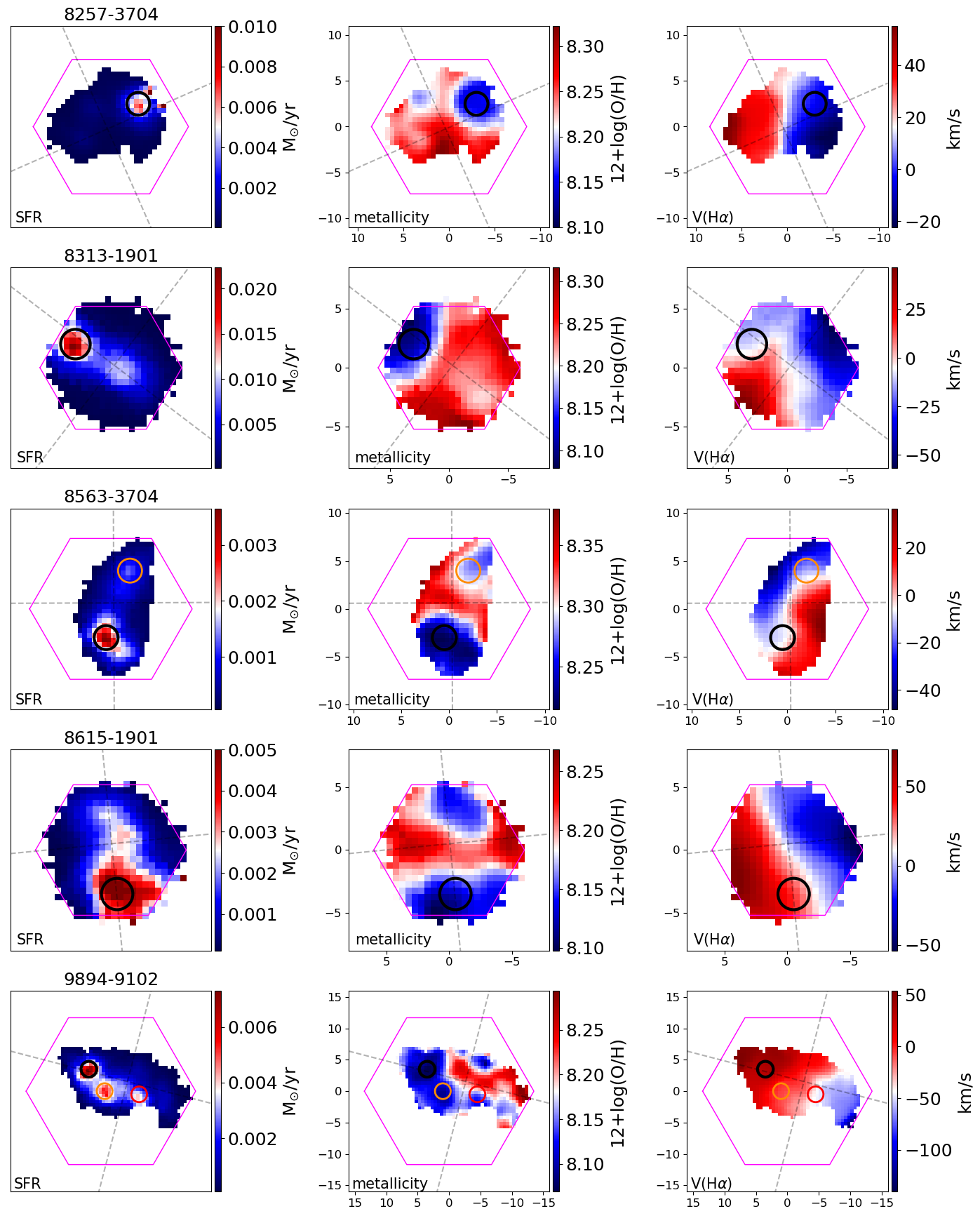} 
    \caption{From left to right: the SFR maps, the metallicity maps, and the H$\alpha$ velocity maps. The black circles, orange circles, and red circles with diameters of 2.5\arcsec\ mark the clumps in the five star-bursting galaxies.}
    \label{fig:maps}
\end{figure*}

With the emission-line flux, we estimate the SFR and gas-phase metallicity of the five star-bursting galaxies. We select the spaxels whose S/N radios of H$\alpha$ emission line are higher than 15, and we find nearly all of these spaxels fall into the star-forming region in the Baldwin-Phillips-Terlevich (BPT) diagram \citep[\nii/H$\alpha$ vs.  \oiii/H$\beta$;][]{BPT1981, kewley2001, kauffmann2003a}.
SFR is calculated with the H$\alpha$ luminosity following \citep{hao2011, kennicutt2012} 

\begin{small}
\begin{equation}
    \log\left(\frac{\text{SFR}}{\text{M}_{\odot}/\text{yr}}\right) = \log\left(\frac{L_{\text{H}\alpha}}{\text{erg/s}}\right)-41.27
    \label{eq:sfr}
\end{equation}
\end{small}

The SFR maps are shown in the first column of Figure~\ref{fig:maps}. There are clumps of high star formations in all five galaxies. Some have multiple off-centered clumps. Based on the peak positions of the clumps identified in the SFR maps, we label them with black, orange, and red circles. These circles are centered at the locations with the highest SFR within each clump, and their diameters are 2.5\arcsec\ . For the galaxies with multiple clumps, the central SFR decreases from black circles to orange circles to red circles.

There are a lot of methods to calculate the oxygen abundance \citep{pagel1979, Storchi1994, Denicolo2002, pp04, tremonti2004, henry2013, marino2014}. In this work, we estimate the metallicity of the spaxels using the calibration obtained for the O3N2 indicator by \cite{marino2014} (Equation~\ref{eq:M13O3N2}). The O3N2 indicator depends on two strong emission line ratios \citep{Alloin1979} (Equation~\ref{eq:O3N2}).

\begin{small}
\begin{equation}
   12+\mathrm{log(O/H)}=8.533[\pm0.012] - 0.214[\pm0.012] \times \mathrm{O3N2}, 
   \label{eq:M13O3N2}
\end{equation} 
\end{small}

where

\begin{small}
\begin{equation}
    O3N2 = \log\frac{\oiii\lambda5007/H\beta}{\nii\lambda6584/H\alpha} 
   \label{eq:O3N2}
\end{equation} 
\end{small}

The second column of Figure~\ref{fig:maps} shows the metallicity maps. In general, the clumps show poor metallicities, which are lower by about 0.1 dex than those of the host galaxies. In galaxies with multiple clumps, there is an inverse relationship between SFR and metallicity, where higher SFR corresponds to lower metallicity. There are some exceptions; for example, 8615-1901 has two low-metallicity regions, with one of them not exhibiting strong SFR.

The gas velocity maps can help us determine if there is any disturbance in the clump regions compared to the gas in the host galaxy.
We obtain the observed H$\alpha$ velocity maps from the Data Analysis Pipeline \citep[{\tt{DAP}};][]{westfall2019, Belfiore2019}, and plot them in the third column of Figure~\ref{fig:maps}. We find that generally these galaxies have rotation-dominated velocity fields. 8313-1901, 8563-3704, and 8615-1901 have clear disturbances in the position of the clumps, indicating that the clumps in them might have external origins. 

\subsection{$g$-band Images in Different Age Intervals}
\label{sec:youngold}

\begin{figure*}
    \centering
    \includegraphics[width=1\textwidth,clip=true,trim=0 0 0 0]{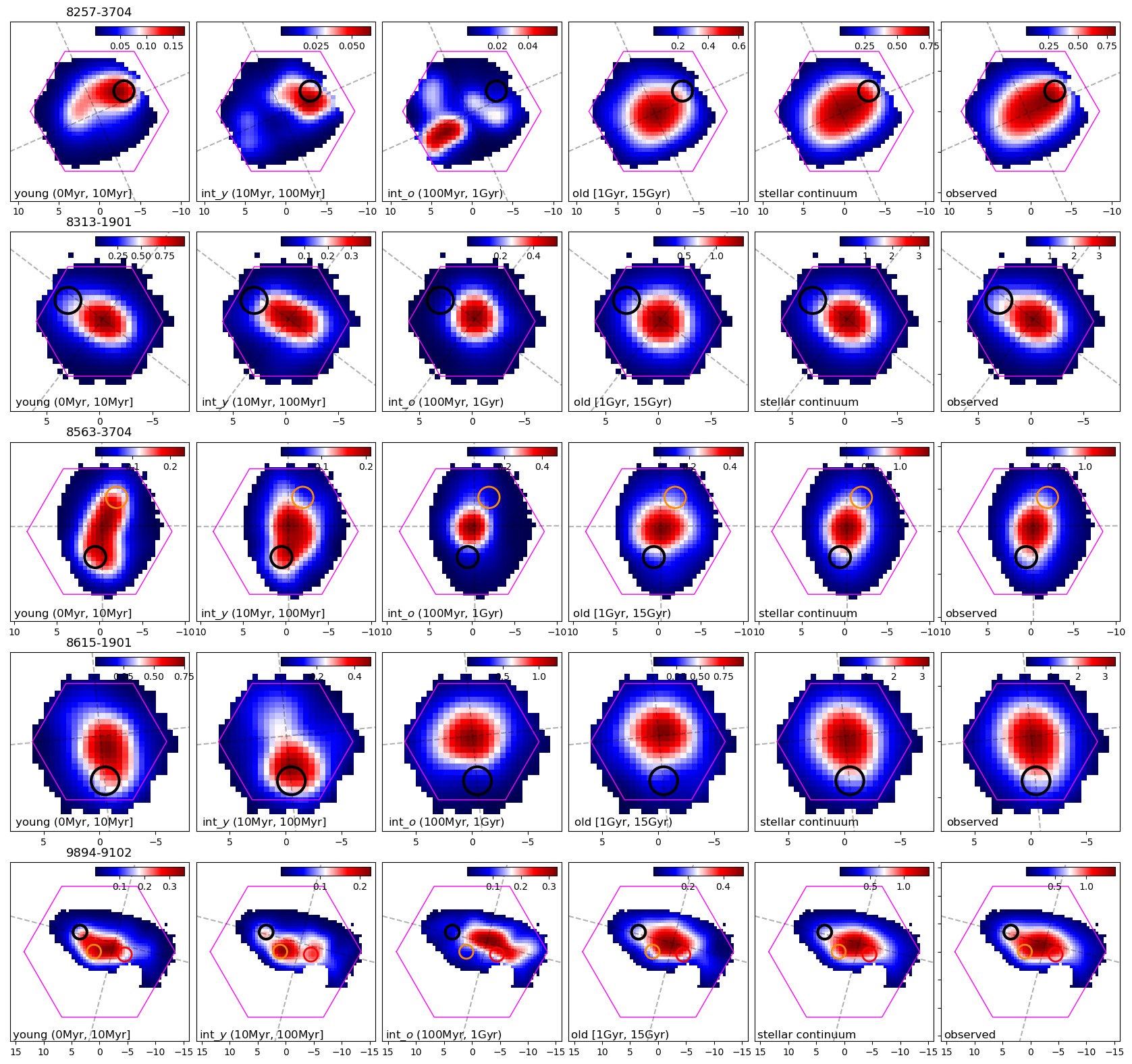}
    \caption{The smoothed $g$-band images of the five galaxies. The unit is nanomaggies. From left to right are the images of young stellar populations (0-10~Myr), intermediate-young stellar populations (10~Myr-100~Myr), intermediate-old stellar populations (100~Myr-1~Gyr), old stellar populations ($\geq$1~Gyr), along with the stellar images and the observed images. }
    \label{fig:gbands}
\end{figure*}

The {\tt FADO} fitting analysis of the IFS data provides spatially-resolved stellar populations. This enables us, in principle, to reconstruct the distribution of stellar fluxes (i.e., the images) at different ages, allowing us to probe how the morphologies of the galaxies can change as a function of evolutionary time \citep{Peterken2019}. We attempt to conduct this investigation in this subsection.

Based on the spatially-resolved SFH and $E(B-V)$ values obtained by {\tt FADO}, we use 90 SSPs to model the attenuated stellar continua for four age intervals at each spaxel: 0-10~Myr (young), 10~Myr-100~Myr ( intermediate young), 100~Myr-1~Gyr (intermediate old ), and $\geq$1~Gyr (old). After convolving the $g$-band filter with the Python code {\tt sedpy} \citep{sedpy} in unit of nanomaggies\footnote{ $\sim$ 3.631 $\mu$Jy; see  https://www.sdss.org/dr17/algorithms/magnitudes/}. We get the $g$-band images of five galaxies in four age intervals, and show in Figure~\ref{fig:gbands}. All $g$-band seeing-limited images are smoothed using a Gaussian profile with FWHM$=$2.5\arcsec for uniform resolution and the positions of the clumps are marked with circles, as in Figure~\ref{fig:maps}. The first to the fourth columns are the reconstructed $g$-band images of young stellar populations, intermediate-young stellar populations, intermediate-old stellar populations, and old stellar populations, respectively. We also calculate the $g$-band stellar images based on the attenuated stellar continua predicted by {\tt FADO} and show them in the fifth column of Figure~\ref{fig:gbands}. The observed images derived from the Cobs spectra are shown in the sixth column for comparison. Like the previous images from the four age intervals, both the stellar images and the observed images are smoothed.

The structures of the $g$-band images of the five galaxies in different age intervals are clearly different from each other. In particular, the images of young, intermediate-young, and intermediate-old stellar populations are more clumpy and asymmetric compared to the images of old stellar populations. This implies that the clumps in the five star-bursting dwarf galaxies only appeared within the past few hundred million years. This finding is in agreement with previous studies \citep{Tosi2009, McQuinn2010}. The morphology of the images of old stellar populations is similar to that of the stellar images, primarily because the $g$-band flux of the old stellar populations is generally greater than that of the younger stellar populations in most galaxies. Below, we provide a detailed description of these images for each galaxy.

\begin{itemize}
    \item 
    8257-3704: 
    An off-centered clump with high SFR and low metallicity is shown to the northwest in the $gri$ image in the first column of Figure~\ref{fig:image}. This clump is visible in both the images of young and intermediate-young stellar populations but not in other age intervals. Additionally, we have identified a clump in the southeastern part of the galaxy in the image of intermediate-old stellar populations.
    The southeastern clump exhibits slightly lower metallicity compared to the host galaxy, although it is not clearly distinguishable in the $gri$ composite image and the SFR map. The image of old stellar populations is more circular than the overall stellar image, indicating that the flux contribution of the young populations in this galaxy cannot be ignored.
    We speculate that this galaxy may have experienced two star formation events within a few hundred million years, with the first one occurring in the southeast and the second one in the northwest.

    \item 
    8313-1901: 
    A distinguishable clump with high SFR and low metallicity is shown in the northeast direction. The morphology of the image of young stellar populations and the observed image is both irregular, while the shape of the galaxy is more symmetric in other age intervals. Our previous analysis indicates that the structural changes of 8313-1901 mainly result from the gas accretion within the recent 7~Myr \citep{Ju2022}.

    \item
    8563-3704:
    Two clumps are visible in the north and south directions in the $gri$ composite image. They have a high SFR and low metallicities. The gas appears to be kinematically separated from the rotating disk of its host galaxy, as observed in the H$\alpha$ velocity map (the third column of Figure~\ref{fig:maps}). The southern clump is present in both the images of young and intermediate-young stellar populations, while the northern clump is also visible in the image of intermediate-old stellar populations. This suggests that the formation epochs of these two clumps are different, with the northern clump forming first and the southern clump forming subsequently.

    \item 
    8615-1901: 
    In the $gri$ composite image, the off-centered clump extends across almost the entire host galaxy. The bluest region in the southern direction of the galaxy exhibits high SFR and low metallicity. The rotation velocity of the gas is the highest among the five galaxies. The orientation of the clump extension changes across different age intervals. In the image of young stellar populations, the peak is located at the galaxy center and extends towards the south. In the images of intermediate-young stellar populations, the peak is located in the south and extends towards the north. In the image of intermediate-old stellar populations, the peak is at the galaxy center and extends to the east. The morphology of the image of old stellar populations shows symmetry. Both the stellar image and the observed image are asymmetric due to the bluest regions of the galaxy.

    \item 
    9894-9102: 
    In the $gri$ composite image, three clumps are observed from east to west. These clumps have high SFR and low metallicities. From east to west, the intensity of star formation in these clumps increases, while the metallicity decreases. Their sequential appearance in the images of intermediate-old, intermediate-young stellar populations, and young stellar populations suggests that these clumps might have formed sequentially from east to west over hundreds of millions of years.

\end{itemize}

\section{Discussion}
\label{sec:discussion}

\subsection{The Properties of The Hosts as They Evolve}
\label{sec:hosts}

\begin{figure}
    \centering   
    \includegraphics[width=0.5\textwidth,clip,trim={0 0 0 0}]{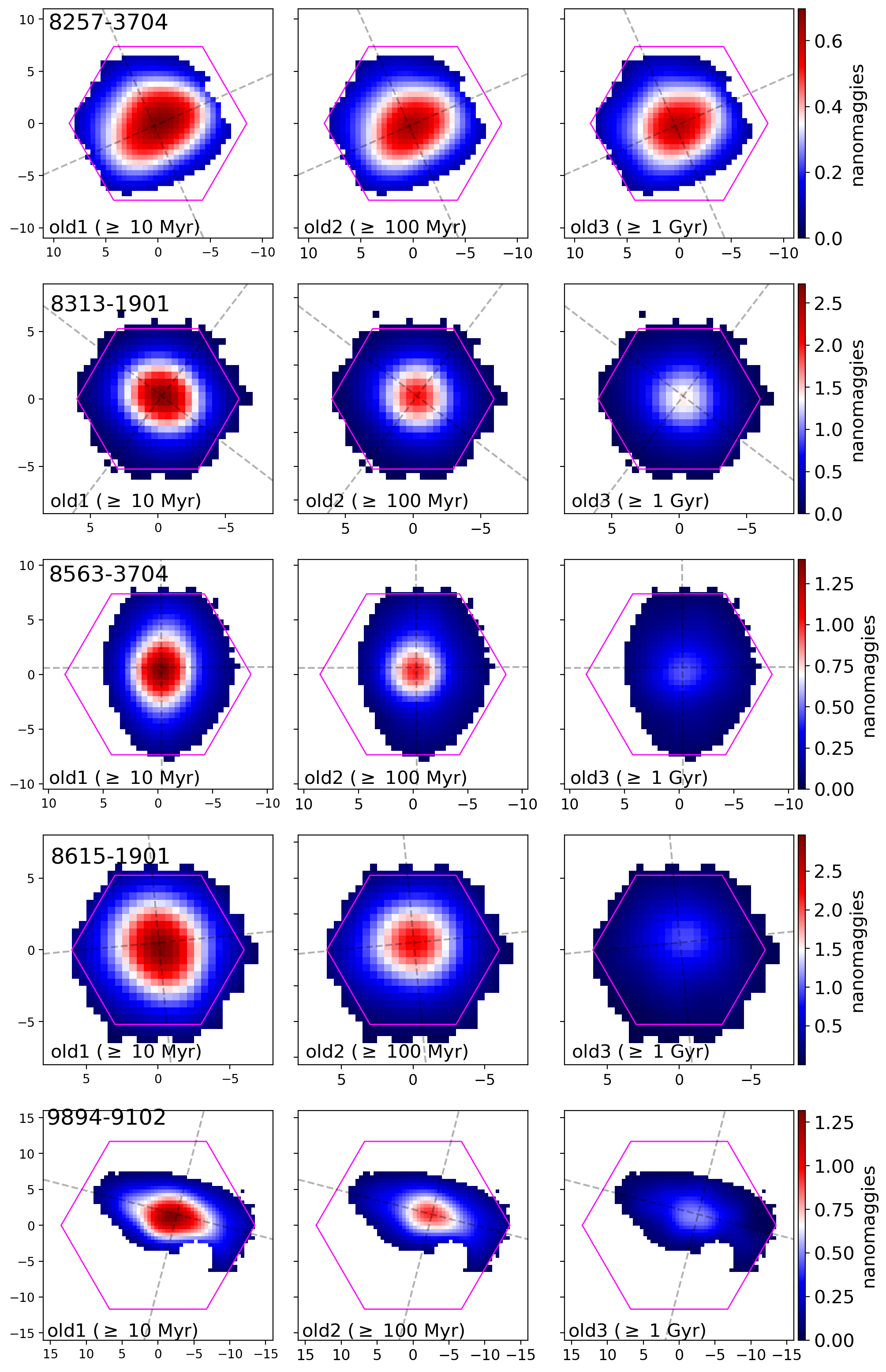} 
    \caption{From left to right columns are the old1 images, old2 images, and old3 images.}
    \label{fig:oldimage}
\end{figure}

\begin{figure}
    \centering   
    \includegraphics[width=0.5\textwidth,clip,trim={0 0 0 0}]{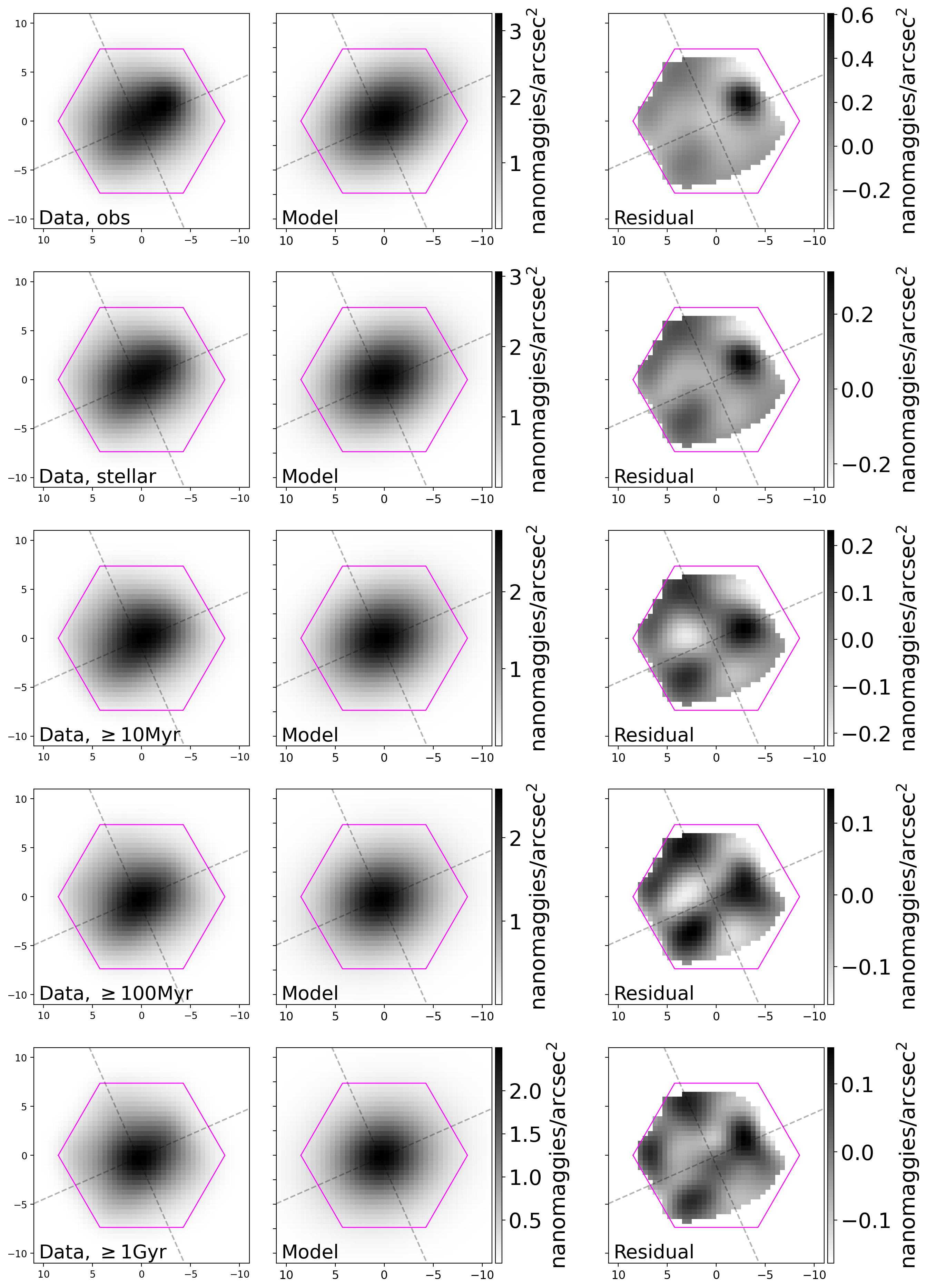} 
    \caption{Two-dimensional surface brightness modeling of 8257-3704 with {\tt GALFIT}. The first column shows the smoothed $g$-band images from the Cobs spectra and stellar continuum, as well as the old1, old2, and old3 images. The second column shows the best {\tt GALFIT} fitting model images. The third column shows the residual images.}
    \label{fig:galfit}
\end{figure}

\begin{figure*}
    \centering
    \includegraphics[width=1.0\textwidth,clip,trim={0 0 0 0}]{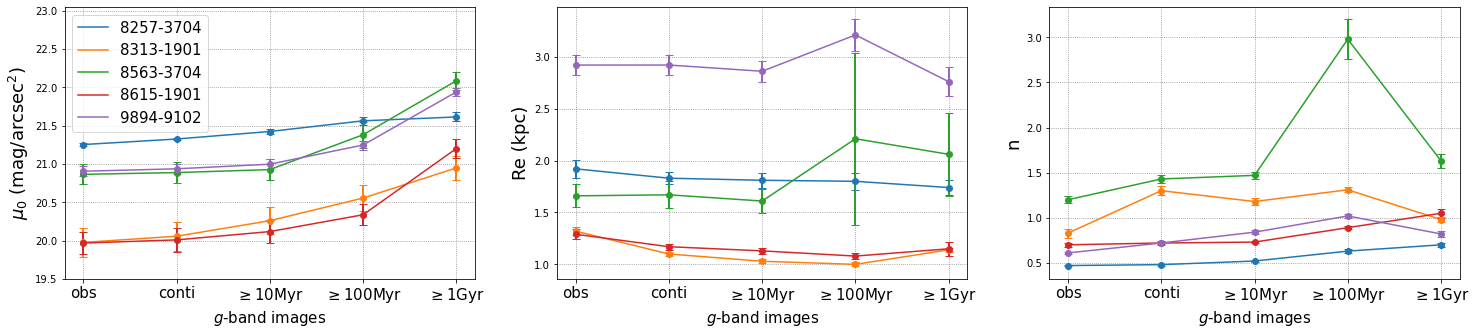} 
    \caption{
    The variation of the structural parameters (central surface brightness $\mu_0$, left; effective radius $R_e$, middle; and S\'ersic index $n$., right) in the image sequence of the observed spectra, the stellar continuum, as well as the old1, old2, and old3 images.}     \label{fig:evolution}
\end{figure*}

\begin{figure}
    \centering    
    \includegraphics[width=0.5\textwidth,clip,trim={0 0 0 0}]{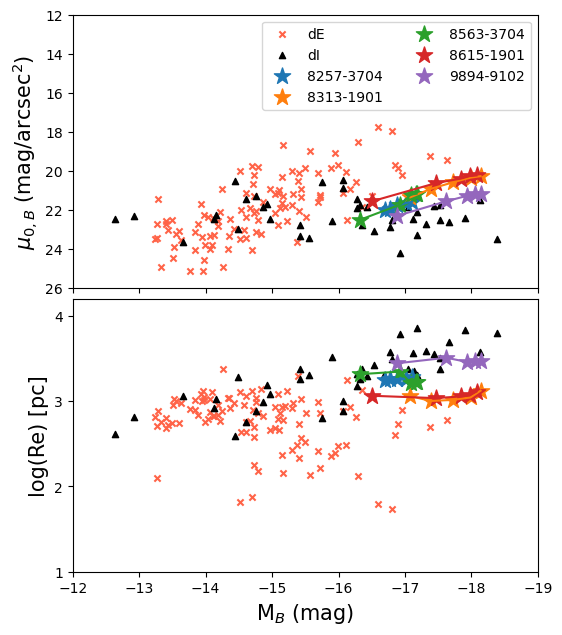} 
    \caption{The central surface brightness ($\mu_0$, the upper panel) and effective radius ($R_e$, the bottom panel) versus the $B$-band absolute magnitudes of galaxies of different types. The orange crosses and black triangles represent the dEs \citep{Binggeli1993, Graham2003} and dIrrs \citep{zee2000}, respectively. The stars in different colors are the five star-bursting galaxies studied in this work. 
    Each galaxy is plotted as five stars that indicate their different $B$-band absolute magnitudes. The brightest star corresponds to the observed spectra, followed by the stellar continuum, the old1 image, the old2 image, and the faintest star represents the old3 image.}
    \label{fig:dEdI}
\end{figure}

In Figure~\ref{fig:gbands}, we find that there are clear variances in the morphology among images of different stellar populations of the five star-bursting dwarf galaxies. Star-forming clumps only ``appear" in younger stellar populations. These hint us that we may be able to use the images of the different stellar ages to trace how these galaxies evolve. In particular, one of the big interests people have on these dwarf galaxies is to find out whether they could be evolutionary connected to other types of dwarf galaxies if they had not been star forming. Previously this can be done on the optical images by masking out the star-forming regions and considering what is left as hosts. Now we can look back at the time before the star formation happened, thereby making use of the full image of the galaxy.

In addition, there may be multiple epochs of star formation. The star-forming clumps that appeared earlier may evolve and eventually become a part of the host galaxies. Therefore, we can consider several stages of the ``hosts". We select three age nodes (10~Myr, 100~Myr, and 1~Gyr) as the representative formation epochs of star formation, then we can construct the images with stellar populations larger than these three age nodes all as the host galaxies, and name them ``old1" ($\geq$10~Myr), ``old2" ($\geq$100~Myr), and ``old3" ($\geq$1~Gyr) separately.

The $g$-band images of ``old1"(left), ``old2"(middle), and ``old3"(right), which are smoothed by a Gaussian profile with an FWHM of 2.5\arcsec, are shown in Figure~\ref{fig:oldimage}. These images are almost all smooth and asymmetric. We use {\tt GALFIT} to obtain the structure of the images \citep{peng2002, peng2010}. This software models the observed light distribution of a galaxy with a best-fit model image convolved with a PSF function. We use a single S\'ersic profile to fit the observed images, the stellar images, the old1 images, the old2 images, and the old3 images. In Figure~\ref{fig:galfit}, we display the fitting results for 8257-3704. The first column comprises the observed image, followed by the stellar image, the old1 image, the old2 image, and the old3 image. The second column shows the model images, while the third column shows the residuals. The fitting results for the remaining four galaxies can be found in Appendix A. The major and minor axes of the galaxies are depicted as gray dashed lines in the images in this work, which are determined based on the structural parameters derived from the old2 images.

In Figure \ref{fig:evolution}, we show the central surface brightness ($\mu_0$), the effective radius ($R_e$), and the S\'ersic index ($n$) of the images of the old1, old2, and old3, stellar images and the observed images. Different lines represent different galaxies. The central surface brightnesses of the five star-bursting dwarf galaxies are generally brighter as they evolve, some can increase for over 1~mag/arcsec$^2$. Comparatively, the effective radius and the S\'ersic index show no significant changes over time. This is probably not as surprising as it looked. One possible reason that we do not observe a clear growth in the size of the disk for these galaxies is that the images are reconstructed from the MaNGA IFU data and have limited spatial resolution. Another more likely reason is related to the fact that the images we construct for different stellar ages are what these stellar populations distribute now, not when they were born. In the early stages of galaxy formation, galaxies tend to exhibit a clumpy morphology due to continuous gas inflow, as studied by \cite{Ceverino2010}. This continuous gas supply maintains a system characterized by a clumpy disk and bulge, which remains in a relatively stable state for several billion years. However, due to various stellar kinematic processes, such as disk rotation and stellar migration, the current appearance of galaxies is smoother than their initial state \citep{2010gfe..book.....M}. As a result, the distribution of stars formed within galaxies may differ from what we observe today. These star-bursting dwarf galaxies might have been more compact than what we currently observe, even for older stellar populations. 


On the other hand, the images of different stellar populations of these star-forming dwarf galaxies as they would have evolved till now, are prefect to be used to compare with other types of dwarf galaxies to probe possible evolutionary connections among them. Previous studies fitted the surface brightness profile of the host galaxies and then compared the $B$-band central surface brightness ($\mu_{0,B}$) and the effective radius ($R_e$) with their absolute $B$-band magnitudes.\citep[e.g.][]{Amorin2009, Lian2015}.
In Figure \ref{fig:dEdI}, we construct a similar diagram and compare the structural parameters of these five galaxies with dwarf elliptical galaxies (dEs) and dwarf irregular galaxies (dIrrs) in different absolute magnitude bins in the $B$-band. We adopt parameters of dEs from \cite{Binggeli1993, Graham2003} (orange crosses) and dwarf dIrrs from \cite{zee2000} (black triangles). The central surface brightness and total magnitudes in the $r$-band for the five star-bursting galaxies are also measured. The magnitudes in $r$-band are converted into those in $B$-band following \cite{Smith2002}. The five star-bursting galaxies are shown as stars, each in a different color. For a given galaxy, the absolute magnitude values in increasing order are obtained from the observed image, the stellar image, the old1 image, the old2 image, and the old3 image. The central brightness ($\mu_{0,B}$, the upper panel) and the effective radius ($R_e$, the bottom panel) of the old images of the five star-bursting dwarfs fall between those of dEs and dIrrs. 9894-9102 (purple stars) appears to be closer to dIrrs, while 8313-1901 (orange stars) and 8615-1901 (red stars) are more similar to dEs in terms of their central surface brightness and effective radii. In our previous work \citepalias{Ju2022}, we proposed that 8313-1901 experienced gas accretion around 7~Myr ago. Combined with the information we obtain here, we speculate that 8313-1901 might have initially been a dwarf elliptical galaxy before the gas accretion event.

\subsection{The Clump Evolution in 8257-3704}
\label{sec:smoothness}

\begin{figure*}
    \centering
    \includegraphics[width=0.95\textwidth,clip,trim={0 0 0 0}]{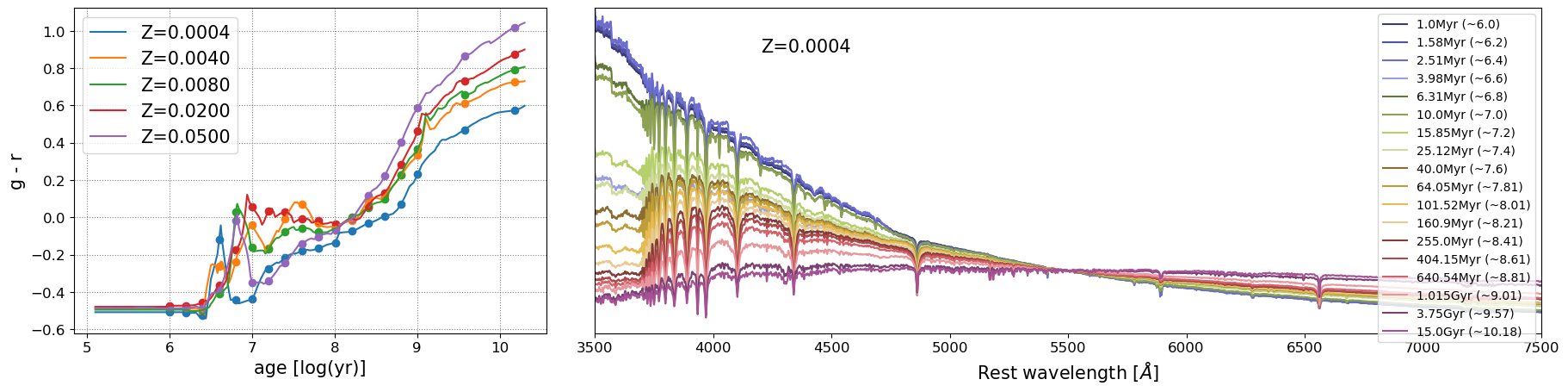} 
    \caption{Top: The $g-r$ color of SSPs in the BC03 library. Dots mark the eighteen ages we used in this work. Bottom: The spectra of the SSPs we used, represented by the stellar metal abundance of 0.0004.}
    \label{fig:ssps}
\end{figure*}

\begin{figure*}
    \centering
    \includegraphics[width=1.0\textwidth,clip,trim={0 0 0 0}]{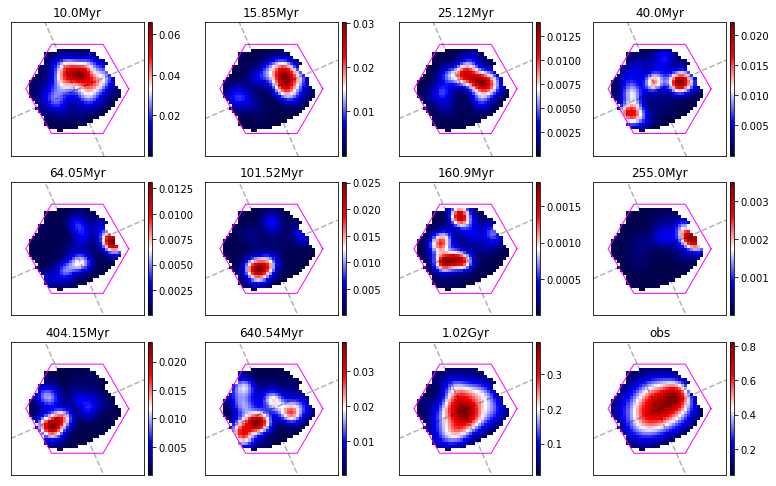} 
   \caption{The smoothed $g$-band images of stellar populations of 8257-3704 at 11 different age intervals, as well as the observed image (bottom-right panel). The units in these images are all nanomaggies. }
    \label{fig:8257}
\end{figure*}

 \begin{figure*}
\centering
\includegraphics[width=1.0\textwidth,clip,trim={0 0 0 0}]{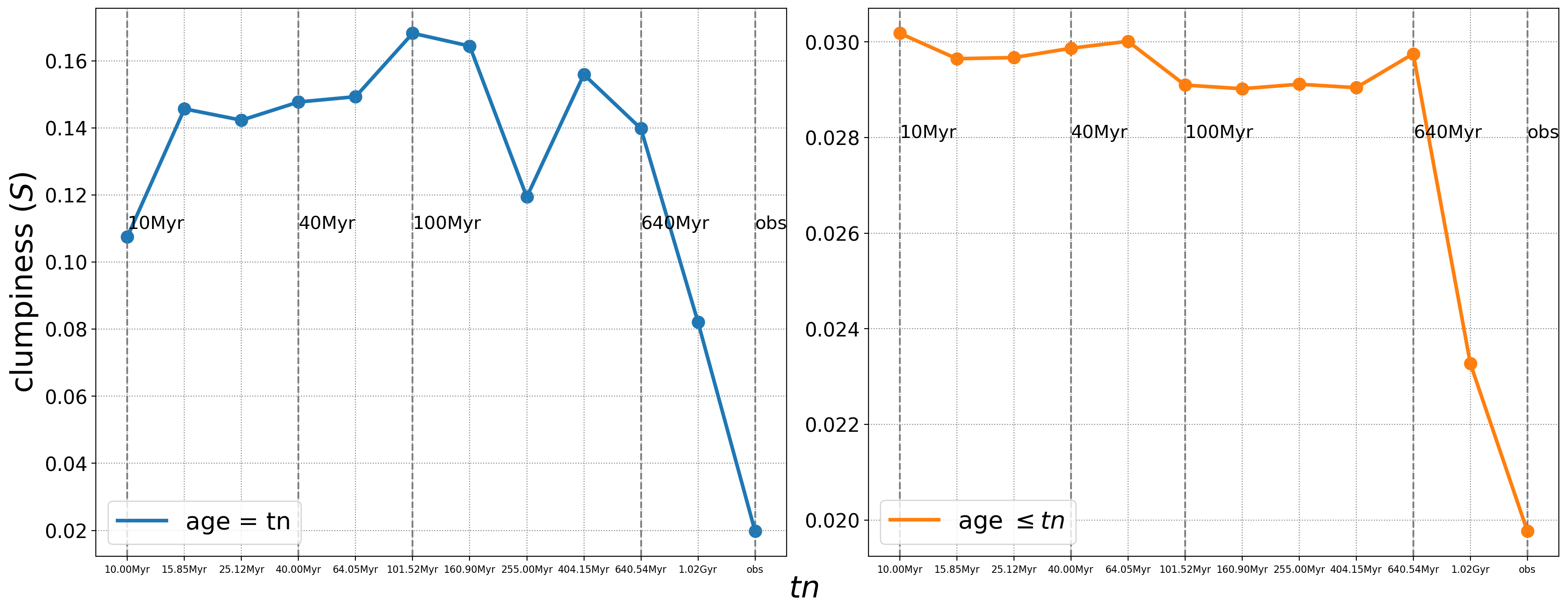} 
   \caption{The clumpiness ($S$) of the $g$-band images of stellar populations across 11 age nodes (age $=$ tn, left panel) and younger than these age nodes (age $\leq$ tn, right panel), with various thresholds of ages. The $S$ of the observed image is shown as the rightmost data point for each panel.}
    \label{fig:smooth}
\end{figure*}

\begin{table}
  \caption{Checking the age robustness using {\tt FADO}}
  \tabcolsep=0.16cm
  \label{table:age}
  \centering
  \begin{tabular}{l c c  }
\hline\hline
 central age & bin width &  FADO age  \\
 Myr & Myr  & Myr \\
\hline
10.00 & 8.00 - 12.00    &	9.63 $\ \pm\ $ 0.00	\\
15.85 & 12.68 - 19.02    &	15.29 $\ \pm\ $ 0.00	\\
25.12 & 20.09 - 30.15	&	25.05 $\ \pm\ $ 0.00	\\
40.00 & 32.00 - 48.00	&	40.00 $\ \pm\ $ 0.00	\\
64.05 & 51.24 - 76.86	&	64.03 $\ \pm\ $ 0.00	\\
101.52 & 81.21 - 121.83	&	118.10 $\ \pm\ $ 0.01	\\
160.90 & 128.72 - 193.08	&	160.90 $\ \pm\ $ 0.00	\\
255.00 & 204.00 - 306.00	&	264.40 $\ \pm\ $ 0.00	\\
404.15 & 323.32 - 282.98	&	404.00 $\ \pm\ $ 0.00	\\
640.54 & 512.43 - 768.65	&	640.50 $\ \pm\ $ 0.00	\\
1015.19	& 812.15 - 1218.23 &	1015.00 $\ \pm\ $ 0.00	\\
\hline
20.00 & 16.00 - 24.00    &	20.33 $\ \pm\ $ 0.00	\\
50.00 & 40.00 - 60.00    &	53.13 $\ \pm\ $ 0.00	\\
500.00 & 400.00 - 600.00	&	545.10 $\ \pm\ $ 0.02	\\
\hline\hline
10.00 & 5.00 - 15.00	&	10.98 $\ \pm\ $ 0.00	\\
15.85 & 7.93 - 23.77	&	16.84 $\ \pm\ $ 0.00	\\
25.12 & 12.55 - 37.68	&	25.07 $\ \pm\ $ 0.00	\\
40.00 & 20.00 - 60.00	&	39.43 $\ \pm\ $ 0.00	\\
64.05 & 32.02 - 96.07	&	58.45 $\ \pm\ $ 0.00	\\
101.52 & 50.76 - 152.28	&	113.60 $\ \pm\ $ 0.02	\\
160.90 & 80.45 - 241.35	&	165.10 $\ \pm\ $ 0.00	\\
255.00 & 127.50 - 382.50	&	263.00 $\ \pm\ $ 0.00	\\
404.15 & 202.07 - 606.22	&	411.00 $\ \pm\ $ 0.03	\\
640.54 & 320.27 - 960.81	&	659.40 $\ \pm\ $ 0.06	\\
1015.19 & 507.59 - 1522.78	&	1659.00 $\ \pm\ $ 1.40	\\
\hline
20.00 & 10.00 - 30.00    &	19.90 $\ \pm\ $ 0.00	\\
50.00 & 25.00 - 75.00    &	59.50 $\ \pm\ $ 0.00	\\
500.00 & 250.00 - 750.00	&	506.70 $\ \pm\ $ 0.01	\\
\hline\hline
\end{tabular}\\
\end{table}

Figure~\ref{fig:gbands} illustrates that off-centered clumps are generally present in the images of the young, intermediate-young, and intermediate-old stellar populations. Considering that these images reflect what the corresponding stellar populations distribute now, not when they were born, the fact that these star-forming clumps remain visible for several hundred million years suggests that the clumps are relatively stable structures and do not easily disintegrate and become part of the disk within a billion years. This suggests that the spatially resolved images of different stellar populations are suitable for studying the properties of the clumps, at least for the last billion years. Among the five galaxies, 8257-3704 stands out as it exhibits two different off-centered clumps in different locations. In the intermediate-old image, the clump is situated in the southeast direction, while in the intermediate-young and young images, it appears in the northwest direction. The southeastern clump is not clear in the $gri$ image. This suggests that 8257-3704 underwent two distinct starburst events: the first one occurred in the southeast direction between 100~Myr and 1~Gyr ago, it seems to have diminished in the 10-100Myr. The second starburst took place in the northwest direction and likely occurred within the past 100~Myr. In this subsection, we attempt to check the stellar images of more refined age intervals to further probe when the southeastern clump might have emerged and vanished. 

We choose to use the 18 age nodes in the 90 SSPs that we used in this work. To double-check whether this set of age sampling is appropriate, in Figure~\ref{fig:ssps}, we plot the $g-r$ colors of the 90 SSPs (left panel) and the spectra of the 18 SSPs with a representative metallicity of Z=0.0004 (right panel). In the left panel, the lines of various colors correspond to different stellar metallicities, while the solid data points denote the 18 ages of the SSPs. In the right panel, the SSPs are normalized to the flux at 5500\,\AA. We observe that SSPs from 10~Myr to 1~Gyr (including 11 age nodes) can be easily distinguished based on their $g-r$ colors and spectral characteristics. Thus, we will mainly do our investigations on the 11 age nodes.


Given the smaller age intervals, we would also like to check that a reasonable result can be obtained by the {\tt FADO} fitting of a spectrum of a certain age without confusing two adjacent age nodes. Therefore, we run a simulation test to examine how {\tt FADO} can reproduce the 11 age nodes from 10~Myr to 1~Gyr fed to {\tt FADO} to fit.

We construct a series of toy-simulated spectra featuring a continuous star formation rate within specific age intervals. These intervals include the aforementioned 11 age nodes, as well as three ages that were randomly chosen. The simulated spectra are built with 221 SSPs in the BC03 library. These SSPs use the Padova1994 stellar evolution tracks and Chabrier IMF with a metallicity of 0.004. We note that the SSPs used in generating the simulated spectra are different from the ones used in the FADO fitting. In Table \ref{table:age}, we show the central age (C) and bin width (W) in the first column and second column, respectively. Two bin widths regulated by the central age are tested: W = C$\times$(1+0.2) - C$\times$(1-0.2)  and W = C$\times$(1+0.5) - C$\times$(1-0.5). The SFR in this age range is 1 M$_\odot$/yr, and the mass-weight stellar ages of simulated spectra are the center ages.

We employ {\tt FADO} and 90 SSPs to fit the simulated spectra. The best-fit mass-weighted mean ages obtained from {\tt FADO} are presented in the third column of Table \ref{table:age}. Comparing these mass-weighted mean stellar ages with the central ages, we find that the mean ages derived from {\tt FADO} closely match the central ages, differing by no more than 5\% for both bin widths. 
So we think the SSPs used in the {\tt FADO} fitting can distinguish the 11 age nodes we adopt.

we plot the smoothed $g$-band flux maps of stellar populations across 11 age nodes (age $=$ tn), ranging from 10~Myr to 1~Gyr in Figure~\ref{fig:8257}. The $g$-band images from $\sim$100~Myr to $\sim$640~Myr reveal a distinct clump located in the southeastern region, while the images of the stellar populations younger than 40~Myr clearly present the northwest clump. We also generate a series of $g$-band images for intervals younger than these age nodes (age $\leq$ tn). 
To quantify the clumpy structures observed in these images, we calculate the clumpiness ($S$) parameter, which is a component of the CAS (Concentration $C$, Asymmetry $A$, and Clumpiness $S$) parameters. The CAS parameters are used for assessing structures and morphology of galaxies \citep{Conselice2003, Lotz2004}.
Clumpiness (also called smoothness) is used to describe the distribution of substructures in galaxy images as defined in Equation \ref{eq:smooth}. 
\begin{equation}
    S = \frac{ {\textstyle \sum_{i,j}^{}\left | I(i,j)- I_{S}(i,j) \right | } }{\textstyle \sum_{i,j}^{}\left | I(i,j) \right |} ,
    \label{eq:smooth}
\end{equation}
where $I(i,j)$ is the original image. $I_S(i,j)$ is an image that has been smoothed by convolving the original image with a box of a given width. A smaller $S$ value indicates fewer substructures. We evaluate the formation time of the off-centered clump in 8257-3704 by analysing the changes in the clumpiness factor in different age ranges.

The effective radius and the galaxy center of the old2 image are used to calculate $S$. $S$ is summed over the spaxels from 0.25$R_p$ to 2.0$R_p$ in Equation \ref{eq:smooth}. The width of the box is 0.25$R_p$. 
Figure~\ref{fig:smooth} shows the clumpiness of images for stellar populations in 11 different age intervals. The blue line represents the clumpiness of the $g$-band images with age equal to these age nodes (age $=$ tn), and the orange line is the $g$-band images with age younger than these age nodes (age $\leq$ tn). We also check the $S$ of the $g$-band images older than these age nodes (age $\geq$ tn), which have little clumpiness as expected. We find that the clumpiness is mainly greater in images with age = 40.00~Myr to 160.90~Myr, it agrees with the $g$-band images in Figure~\ref{fig:8257} that the southeastern clump and the northwestern clump exist at the same time. The clumpiness of the $g$-band images with age = 404.15~Myr and 640.54~Myr are slightly lower than the 160.90~Myr image, it may caused by that only the southeastern clump formed in these age intervals. The clumpiness of the $g$-band images with age younger than 11 age intervals (the orange line in Figure~\ref{fig:smooth}) shows an inflection around 640~Myr. Summarizing from the above observations, we think the southeastern clump survived nearly from 40~Myr to 600~Gyr and the northwestern clump formed nearly 100~Myr ago.

\section{Summary}
\label{sec:summary}

In this work, we apply the stellar population synthesis tool {\tt FADO} and 90 SSPs to the spatially-resolved spectra in five star-bursting dwarf galaxies, selected from the MaNGA survey to have off-centered clumps. The spatially-resolved star formation history obtained from the analysis allows us to look back the time and study how the clumps and the hosts might have evolved. We find that the images of younger stellar populations of these galaxies are significantly more asymmetric and clumpier than the images of stellar populations older than 1~Gyr. Most of the clumps in the five galaxies appeared around hundreds of millions of years ago.  
In some of the 5 galaxies, there are multiple clumps which appear at different locations and even different ages. 8257-3704 is particularly interesting as its southeastern clump is only visible in the images of the stellar populations of 100~Myr-1~Gyr age, but not in the observed $gri$ image. We experiment with constructing the $g$-band images of the stellar populations of refined age intervals by sampling 11 stellar population ages between the 10~Myr to 1~Gyr. We find that this galaxy may have experienced two significant starburst events. The first one occurred around 600~Myr ago and ended around 40~Myr ago, while the second starburst event occurred within the last 100~Myr.

We also construct images of stellar populations older than certain ages to probe the properties and evolution of these properties of the hosts. These images allow us to probe the evolutionary connections between these star-bursting dwarf galaxies and other types of dwarf, such as dEs and dIrrs, in a novel way that had not been fully explored before. We divide the stellar populations into three age intervals ($\geq$ 10 Myr, $\geq$ 100 Myr, $\geq$ 1 Gyr), trying to capture the galaxies before their significant star-formation events which may occur at different epochs. We use {\tt GALFIT} to fit the surface brightness profiles of these ``host" galaxies and then compare their structural parameters with those of other types of dwarf galaxies. We find that the $B$-band central surface brightness and effective radii of these five galaxies, when plotted against their $B$-band magnitude, mainly fall between the regions of dEs and dIrrs. Among them, 8313-1901 and 8615-1901 are closer in their properties to dEs, while 9894-9102 is closer to dIrrs. We speculate that 8313-1901 was a dwarf elliptical galaxy before it accreted gas and formed its current star-forming clump around 10~Myr ago.

By applying the spectral synthesis methods to the IFU data, we are able to obtain the images of galaxies in different age intervals and spatially resolve the SFH. From these five galaxies, we find that this method allows us to acquire more characteristics and evolutionary history of both the host galaxies and clumps in star-bursting galaxies. This method can be applied to larger samples of galaxies of various types. With observations of higher spatial IFU instrumentation such as the IFS onboard the Chinese Space Station Telescope (CSST), we may also use the method to resolve and analyse the stellar population evolution of the star-forming clumps. These analyses will enhance our understanding of galaxy evolution further in the future.

\begin{acknowledgements}
This work was supported by National Key R\&D Program of China No.2022YFF0503402, the National Natural Science Foundation of China (NSFC) grants (Nos. 12233005 and 12041302). 
J.Y. acknowledges support from the Natural Science Foundation of Shanghai (Project Number: 22ZR1473000) and the Program of Shanghai Academic Research Leader (No. 22XD1404200).
Y.R. acknowledges supports from the CAS Pioneer Hundred Talents Program, USTC Research Funds of the Double First-Class Initiative, as well as the NSFC grant 12273037. J.W. acknowledges the NSFC grants 12033004, 12333003.

This work made use of the High Performance Computing Resource in the Core Facility for Advanced Research Computing at Shanghai Astronomical Observatory.

Funding for the Sloan Digital Sky Survey IV has been provided by the Alfred P.Sloan Foundation, the U.S. Department of Energy Office of Science, and the Participating Institutions. SDSS-IV acknowledges support and resources from the Center for High-Performance Computing at the University of Utah. The SDSS website is www.sdss.org.

SDSS-IV is managed by the Astrophysical Research Consortium for the Participating Institutions of the SDSS Collaboration including the Brazilian Participation Group, the Carnegie Institution for Science, Carnegie Mellon University, Center for Astrophysics — Harvard \& Smithsonian, the Chilean Participation Group, the French Participation Group, Instituto de Astrof\'isica de Canarias, The Johns Hopkins University, Kavli Institute for the Physics and Mathematics of the Universe (IPMU) / University of Tokyo, the Korean Participation Group, Lawrence Berkeley National Laboratory, Leibniz Institut f\"ur Astrophysik Potsdam (AIP), Max-Planck-Institut f\"ur Astronomie (MPIA Heidelberg), Max-Planck-Institut f\"ur Astrophysik (MPA Garching), Max-Planck-Institut f\"ur Extraterrestrische Physik (MPE), National Astronomical Observatories of China, New Mexico State University, New York University, University of Notre Dame, Observat\'ario Nacional / MCTI, The Ohio State University, Pennsylvania State University, Shanghai Astronomical Observatory, United Kingdom Participation Group, Universidad Nacional Aut\'onoma de M\'exico, University of Arizona, University of Colorado Boulder, University of Oxford, University of Portsmouth, University of Utah, University of Virginia, University of Washington, University of Wisconsin, Vanderbilt University, and Yale University.

\end{acknowledgements}

\appendix                  

\section{Fitting Results of Surface Brightness for the Rest Four Galaxies}

\begin{figure*}[ht!]
    \centering   \includegraphics[width=0.9\textwidth,clip,trim={0 0 0 0}]{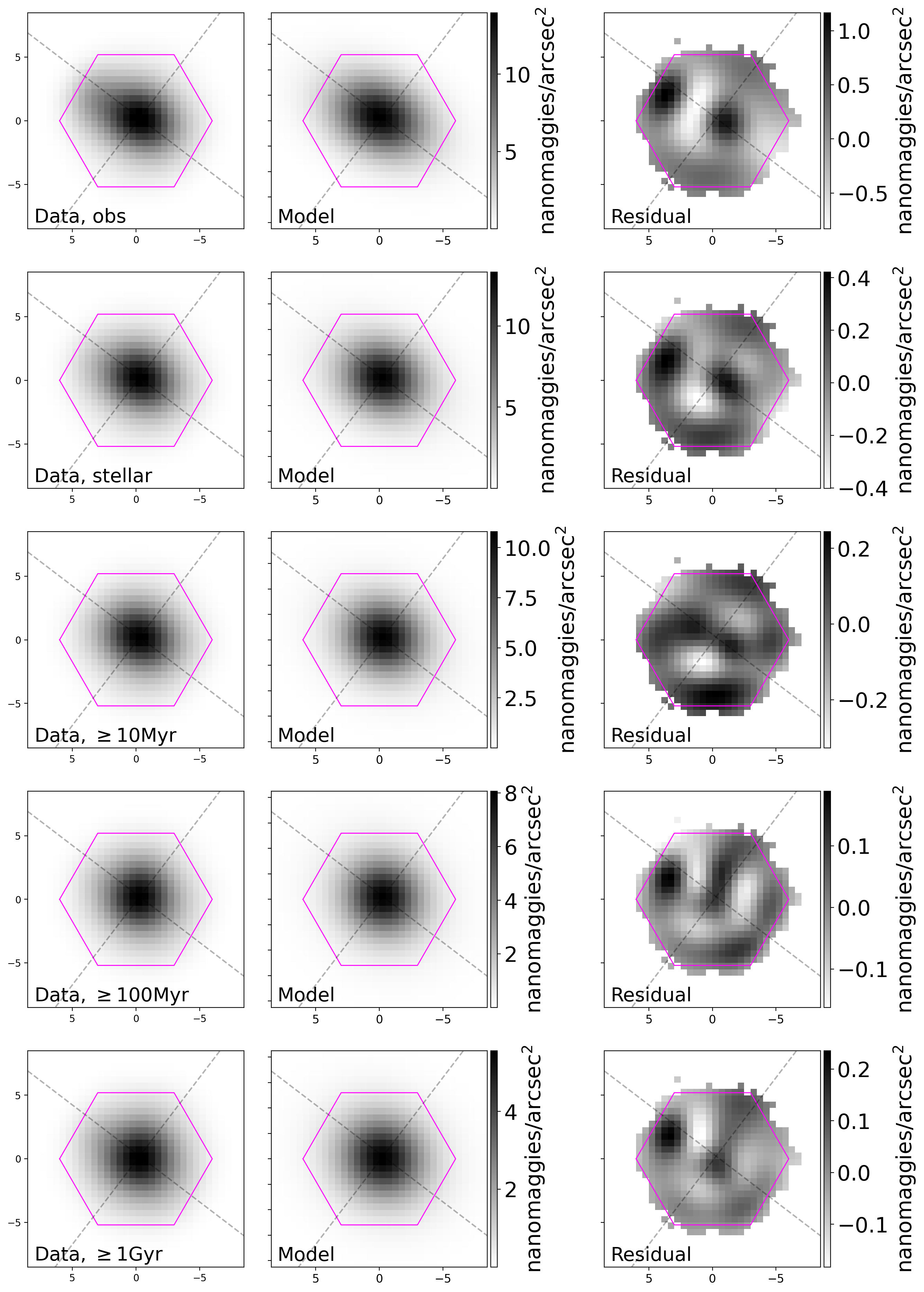} 
    \caption{Two-dimensional surface brightness modeling of 8313-1901 with {\tt GALFIT}.}
    \label{fig:oldimage8313}
\end{figure*}

\begin{figure*}[ht!]
    \centering   \includegraphics[width=0.9\textwidth,clip,trim={0 0 0 0}]{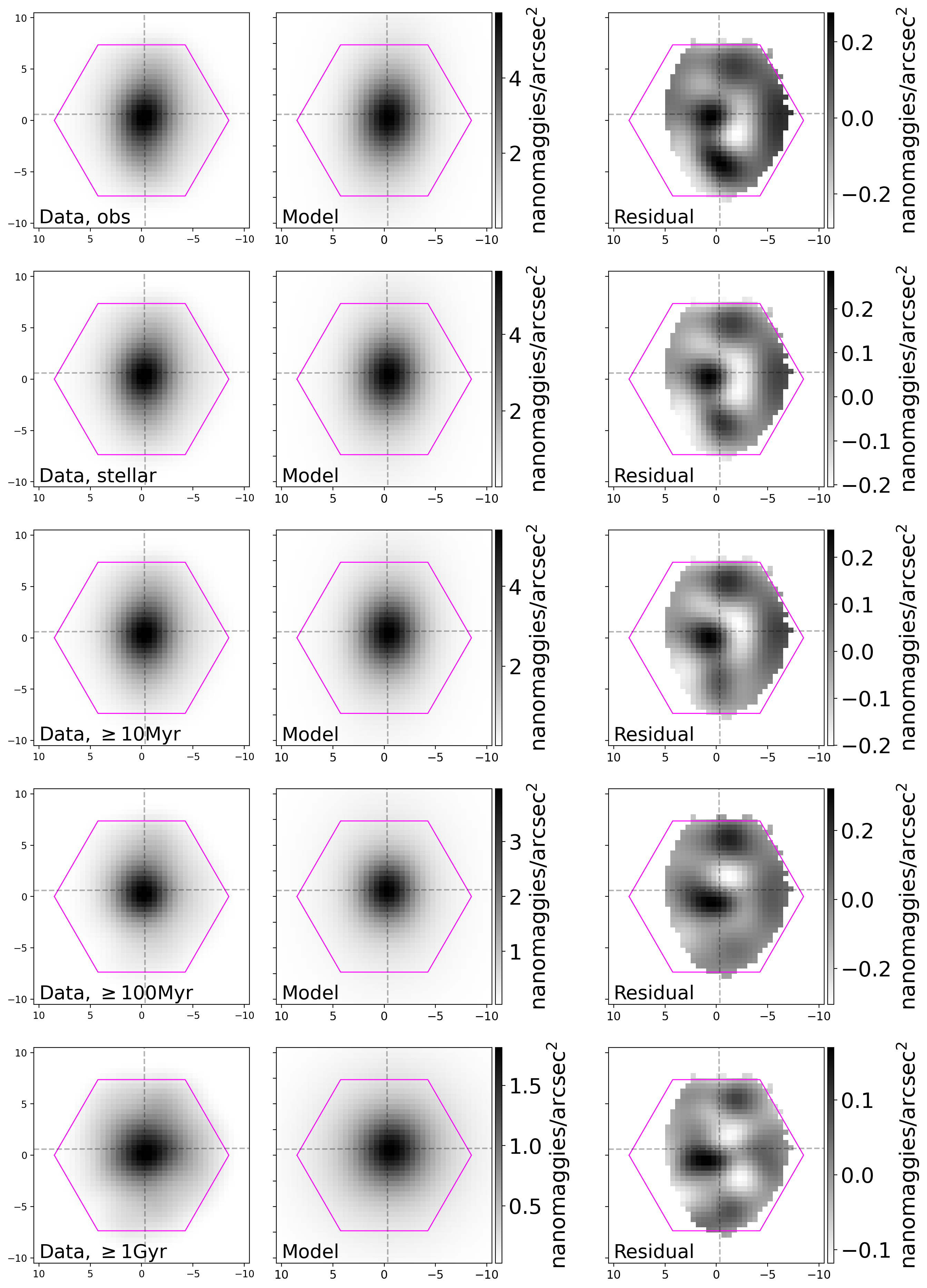} 
    \caption{Two-dimensional surface brightness modeling of 8563-3704 with {\tt GALFIT}. }
    \label{fig:oldimage8563}
\end{figure*}

\begin{figure*}
    \centering   \includegraphics[width=0.9\textwidth,clip,trim={0 0 0 0}]{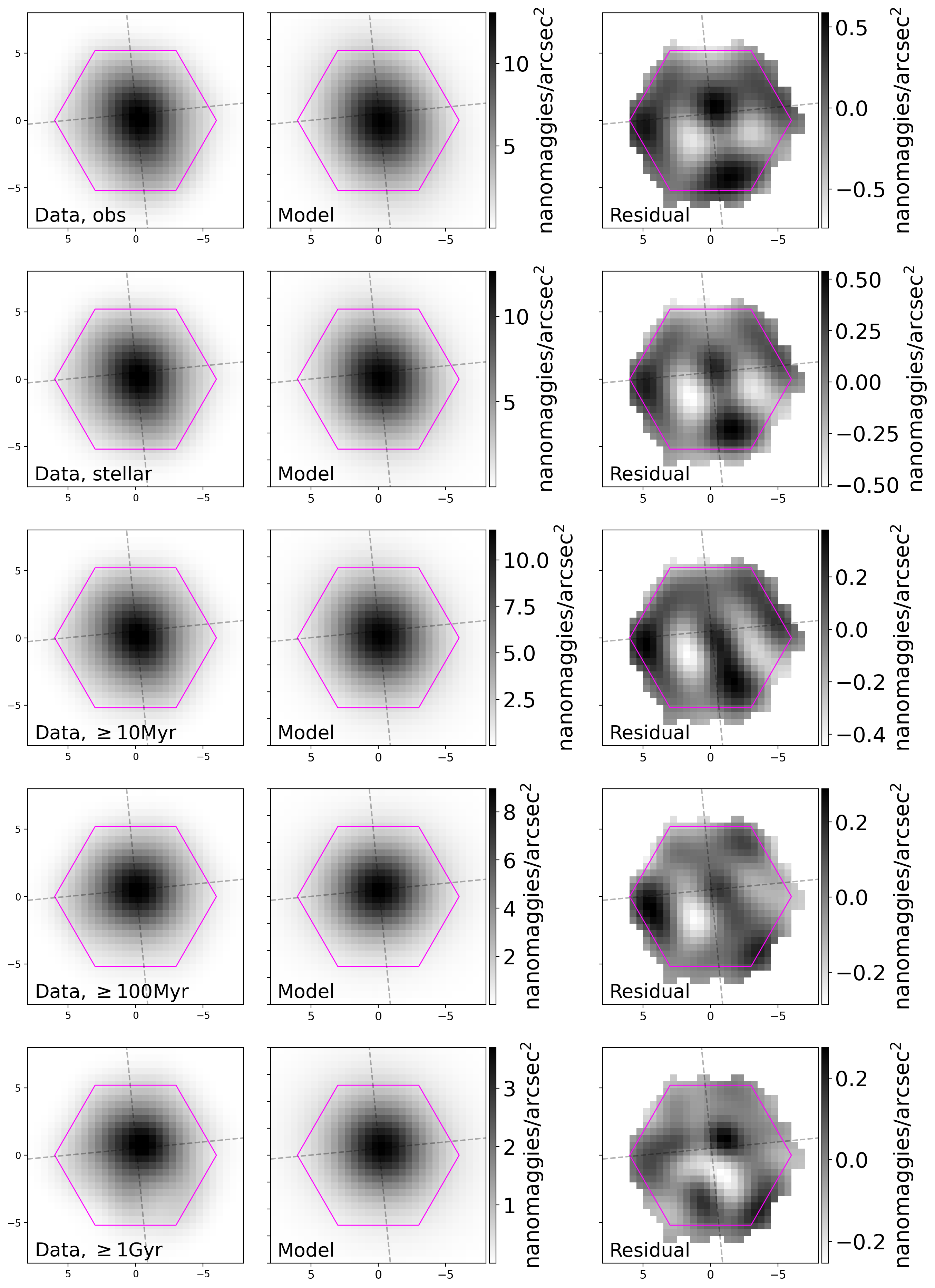} 
    \caption{Two-dimensional surface brightness modeling of 8615-1901 with {\tt GALFIT}. }
    \label{fig:oldimage8615}
\end{figure*}

\begin{figure*}
    \centering   \includegraphics[width=0.9\textwidth,clip,trim={0 0 0 0}]{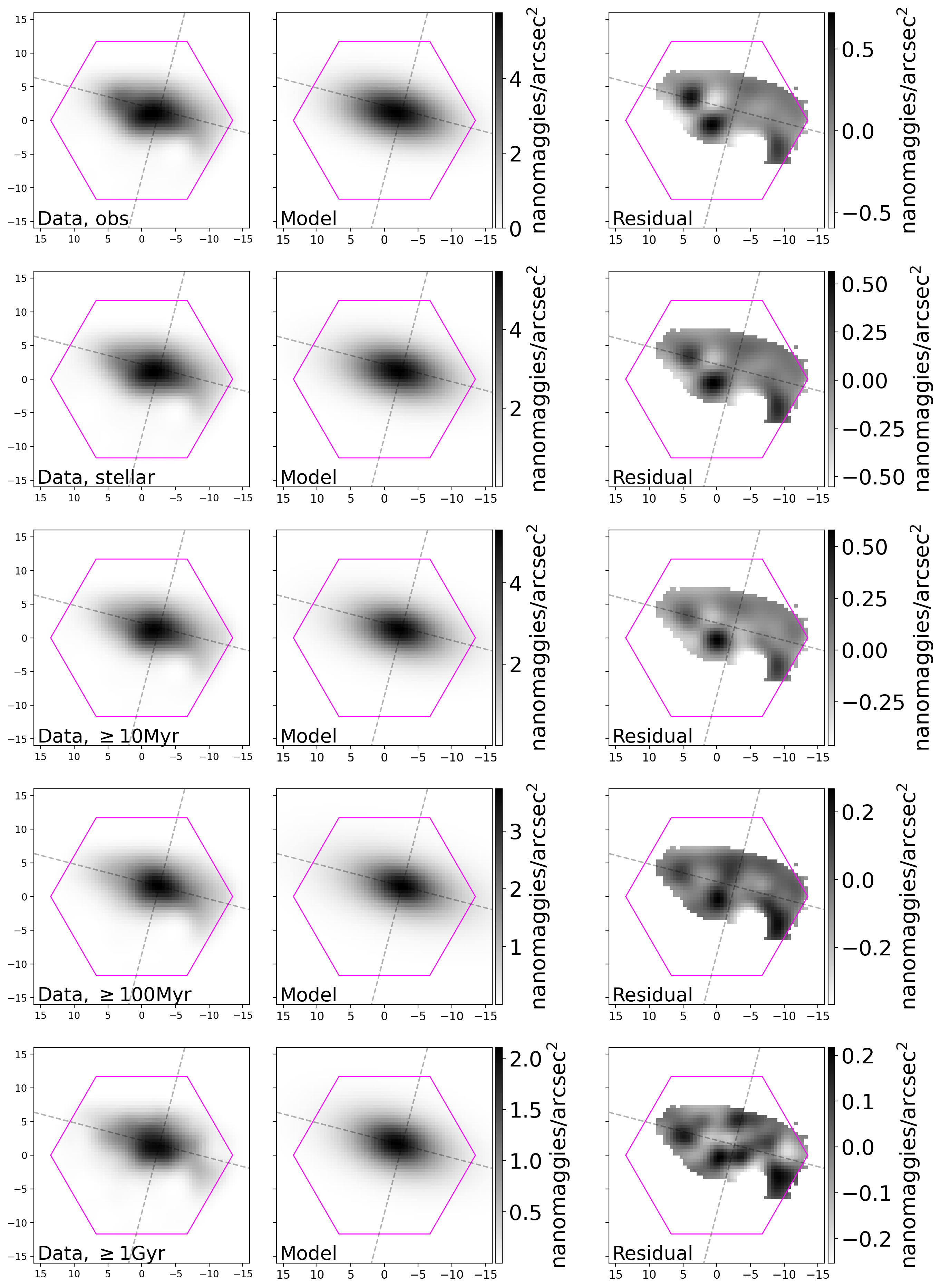} 
    \caption{Two-dimensional surface brightness modeling of 9894-9102 with {\tt GALFIT}. }
    \label{fig:oldimage9894}
\end{figure*}

\begin{table*}
  \caption{The structural parameters of the $g$-band images of the host galaxies.}
  \tabcolsep=0.20cm
  \label{table:2}
  \centering
  \begin{tabular}{l c c c c c c }
\hline\hline
 plateifu  & M$_g$ & R$_e$ (kpc) & n & q & PA & age \\
\hline
 8257-3704	& -17.06$\pm$0.01   & 1.79$\pm$0.06 & 0.70$\pm$0.04 & 0.81$\pm$0.01 & -69.02$\pm$0.84 & old3 ($\geq$1~Gyr)\\
    & -17.13$\pm$0.01   & 1.80$\pm$0.06 & 0.63$\pm$0.04 & 0.76$\pm$0.01 & -66.22$\pm$0.73 & old2 ($\geq$100~Myr)\\
    & -17.23$\pm$0.01   & 1.81$\pm$0.06 & 0.52$\pm$0.03 & 0.73$\pm$0.01 & -64.79$\pm$0.66 & old1 ($\geq$10~Myr)\\
    & -17.33$\pm$0.01   & 1.83$\pm$0.06 & 0.48$\pm$0.03 & 0.70$\pm$0.01 & -61.43$\pm$0.50 & stellar\\
    & -17.58$\pm$0.01   & 1.92$\pm$0.07 & 0.47$\pm$0.03 & 0.70$\pm$0.01 & -58.56$\pm$0.99 & observed\\  
\hline
 8313-1901	& -17.50$\pm$0.03   & 1.14$\pm$0.04 & 0.98$\pm$0.08 & 0.83$\pm$0.01 & 57.48$\pm$0.97 & old3 ($\geq$1~Gyr)\\
    & -17.78$\pm$0.01   & 1.00$\pm$0.04 & 1.31$\pm$0.06 & 0.83$\pm$0.01 & 52.69$\pm$0.86 & old2 ($\geq$100~Myr)\\
    & -18.07$\pm$0.01   & 1.03$\pm$0.04 & 1.18$\pm$0.07 & 0.72$\pm$0.01 & 63.01$\pm$0.58 & old1 ($\geq$10~Myr)\\
    & -18.29$\pm$0.01   & 1.10$\pm$0.04 & 1.30$\pm$0.07 & 0.62$\pm$0.01 & 63.67$\pm$0.48 & stellar\\
    & -18.65$\pm$0.01   & 1.32$\pm$0.04 & 0.83$\pm$0.09 & 0.61$\pm$0.01 & 62.52$\pm$0.57 & observed\\  
\hline
 8563-3704	& -16.74$\pm$0.03   & 2.06$\pm$0.21	& 1.63$\pm$0.06 & 0.95$\pm$0.01 & -7.00$\pm$0.44 & old3 ($\geq$1~Gyr)\\
    & -17.44$\pm$0.05   & 2.21$\pm$0.38	& 2.98$\pm$0.08 & 0.73$\pm$0.01 & -5.19$\pm$0.39 & old2 ($\geq$100~Myr)\\
    & -17.41$\pm$0.01   & 1.61$\pm$0.10	& 1.47$\pm$0.04 & 0.61$\pm$0.02 & -4.69$\pm$0.4 & old1 ($\geq$10~Myr)\\
    & -17.48$\pm$0.01   & 1.67$\pm$0.10 & 1.43$\pm$0.04 & 0.58$\pm$0.01 & 0.23$\pm$1.16 & stellar\\
    & -17.74$\pm$0.01   & 1.66$\pm$0.09 & 1.20$\pm$0.05 & 0.61$\pm$0.01 & -25.4$\pm$6.20 & observed\\  
\hline
 8615-1901	& -16.89$\pm$0.03  & 1.15$\pm$0.07 & 1.05$\pm$0.06	 & 0.89$\pm$0.01 & 58.36$\pm$2.68 & old3 ($\geq$1~Gyr)\\
    & -17.77$\pm$0.02  & 1.08$\pm$0.05 & 0.89$\pm$0.04	 & 0.86$\pm$0.01 & -84.42$\pm$1.06 & old2 ($\geq$100~Myr)\\
    & -18.11$\pm$0.02  & 1.13$\pm$0.05 & 0.73$\pm$0.03	 & 0.87$\pm$0.01 & 33.62$\pm$1.16 & old1 ($\geq$10~Myr)\\
    & -18.22$\pm$0.01 & 1.17$\pm$0.05  & 0.72$\pm$0.04  & 0.82$\pm$0.01 & 27.48$\pm$0.90 & stellar\\
    & -18.26$\pm$0.01  & 1.29$\pm$0.05 & 0.70$\pm$0.04  & 0.81$\pm$0.01 & 25.32$\pm$0.94 & observed\\  
\hline
 9894-9102	& -17.24$\pm$0.04   & 2.76$\pm$0.06 & 0.82$\pm$0.04 & 0.47$\pm$0.01  & 74.49$\pm$0.40 & old3 ($\geq$1~Gyr)\\
    & -17.94$\pm$0.04   & 3.21$\pm$0.06 & 1.02$\pm$0.03 & 0.40$\pm$0.01  & 75.42$\pm$0.20 & old2 ($\geq$100~Myr)\\
    & -18.22$\pm$0.04   & 2.86$\pm$0.05 & 0.84$\pm$0.03 & 0.42$\pm$0.01  & 76.73$\pm$0.22 & old1 ($\geq$10~Myr)\\
    & -18.32$\pm$0.01   & 2.92$\pm$0.05 & 0.72$\pm$0.02 & 0.42$\pm$0.01 & 77.78$\pm$0.23 & stellar\\
    & -18.36$\pm$0.01   & 2.92$\pm$0.05 & 0.61$\pm$0.02 & 0.43$\pm$0.01 & 75.09$\pm$0.23 & observed\\  
\hline
\end{tabular}\\
\end{table*}

From Figure~\ref{fig:oldimage8313} to Figure~\ref{fig:oldimage9894}, we present the $g$-band surface brightness profile fitting results of the host galaxies for 8313-1901, 8563-3704, 8615-1901, and 9894-9102 conducted using {\tt GALFIT}. The first column includes the observed images, the stellar images, old1 images, old2 images, and old3 images. The second column displays the model images given by {\tt GALFIT}, while the third column exhibits the residual images. Table~\ref{table:2} provides the structural parameters, including $g$-band absolute magnitude (M$_g$), effective radii (R$_e$), S\'ersic index ($n$)  axis ratio ($q$) and position angle (PA), for these images as determined by {\tt GALFIT}.

\bibliographystyle{raa}
\bibliography{bibtex}

\end{document}